\documentclass[a4paper,twocolumn,notitlepage,aps,prx,superscriptaddress,10pt]{revtex4-2}
\usepackage{bbm}
\usepackage{amsmath}
\usepackage{amssymb}
\usepackage{amsthm}
\usepackage{amsfonts}
\usepackage{amsmath}
\usepackage{graphicx}
\usepackage{xcolor}
\usepackage{csquotes}
\usepackage{braket}
\usepackage{setspace}
\usepackage{verbatim}
\usepackage[colorlinks=true,linkcolor=teal,citecolor=teal,urlcolor=teal]{hyperref}
\usepackage{mathtools}
\usepackage{booktabs}
\usepackage{braket}
\usepackage{mleftright}
\usepackage{siunitx}
\RequirePackage{snapshot}

\DeclareSIUnit\hartree{Ha}

\theoremstyle{definition}

\begin{document}

\title{Scalable tensor-network error mitigation for near-term quantum computing}

\author{Sergei Filippov}

\affiliation{Algorithmiq Ltd, Kanavakatu 3 C, FI-00160 Helsinki, Finland}

\author{Matea Leahy}

\affiliation{Algorithmiq Ltd, Kanavakatu 3 C, FI-00160 Helsinki, Finland}
\affiliation{Trinity Quantum Alliance, Unit 16, Trinity Technology and Enterprise Centre, Pearse Street, D02 YN67, Dublin 2, Ireland}

\author{Matteo A.~C.~Rossi}

\affiliation{Algorithmiq Ltd, Kanavakatu 3 C, FI-00160 Helsinki, Finland}

\author{Guillermo Garc\'{\i}a-P\'{e}rez}

\affiliation{Algorithmiq Ltd, Kanavakatu 3 C, FI-00160 Helsinki, Finland}

\begin{abstract}
Until fault-tolerance becomes implementable at scale, quantum computing will heavily rely on noise mitigation techniques.
While methods such as zero noise extrapolation with probabilistic error amplification (ZNE-PEA) and probabilistic error cancellation (PEC) have been successfully tested on hardware recently, their scalability to larger circuits may be limited.
Here, we introduce the tensor-network error mitigation (TEM) algorithm, which acts in post-processing to correct the noise-induced errors in estimations of physical observables.
The method consists of the construction of a tensor network representing the inverse of the global noise channel affecting the state of the quantum processor, and the consequent application of the map to informationally complete measurement outcomes obtained from the noisy state.
TEM does therefore not require additional quantum operations other than the implementation of informationally complete POVMs, which can be achieved through randomised local measurements.
The key advantage of TEM is that the measurement overhead is quadratically smaller than in PEC.
We test TEM extensively in numerical simulations in different regimes.
We find that TEM can be applied to circuits of twice the depth compared to what is achievable with PEC under realistic conditions with sparse Pauli-Lindblad noise, such as those in [E. van den Berg \emph{et al.}, Nat.~Phys. (2023)].
By using Clifford circuits, we explore the capabilities of the method in wider and deeper circuits with lower noise levels.
We find that in the case of 100 qubits and depth 100, both PEC and ZNE fail to produce accurate results by using $\sim 10^5$ shots, while TEM succeeds.
\end{abstract}

\maketitle

\section{Introduction} \label{section-introduction}

All roadmaps toward practical quantum computing focus on finding ways to suppress errors and increase the number of logical qubits available. Whereas the long-term goal is to achieve fault-tolerant quantum computing by implementing qubit-demanding error-correcting codes and diminishing the noise below a certain threshold~\cite{calderbank-1996,preskill-1998,gottesman-2022,suzuki-2022}, near-term computing uses all physical qubits as logical ones and significantly relies on the error mitigation techniques compensating the detrimental noise effects in medium-depth quantum circuits~\cite{bharti-2022}. The latter approach attracts increasing attention in view of prospects for advantageous quantum simulations of molecules and binding affinities between chemical compounds \cite{malone-2022,kirsopp-2022,robert-2021} as well as complex quantum dynamics~\cite{keenan-2022,kamakari-2022,plenio-2023,kim-2023}. While mitigating errors is known to require exponential resources in the worst case~\cite{quek-2022}, there is a potential for computational advantage with respect to classical methods if the exponents are low enough, that is, for low enough levels of noise.

Some noise mitigation strategies are agnostic to the nature of the noise, thus providing universality~\cite{rubin-2018,mcardle-2019,koczor-2021,suchsland-2021,czarnik-2021,strikis-2021}. However, the precise knowledge of the noise model generally makes it possible to cancel errors in a more efficient way. Prominent algorithms in this regard are probabilistic error cancellation (PEC)~\cite{temme-2017,berg-2023,piveteau-2022} and zero-noise extrapolation with probabilistic error amplification (ZNE-PEA)~\cite{kim-2023}. PEC represents a noise-free circuit as a quasi-probability distribution of the randomised noisy ones at the expense of a measurement overhead (which quantitatively shows the increase in the measurement outcomes needed to get the same precision in estimation of observables). The ZNE-PEA intentionally increases the strength of the characterised noise by sampling gates from a true probability distribution, thus avoiding the measurement overhead but suffering from a potential bias in the estimated quantities and extrapolation instabilities for large circuit depths. 

The recently reported ZNE-PEA experiment~\cite{kim-2023} was followed (among others) by purely classical approximate tensor network simulations of noiseless quantum processors~\cite{tindall-2023,anand-2023} that qualitatively agree with the experiment but differ quantitatively. Here we show that the rivalry between the quantum hardware and the classical tensor networks can be converted into a fruitful collaboration. An example of this vision is presented in Ref.~\cite{guo-2022}, aiming to find the approximate noise inversion map via tensor network methods and simulate this map via single-qubit gates in quantum hardware. We propose another approach, where the noise mitigation is entirely performed at the classical post-processing stage and the tensor network methods shine as they are not restricted by the requirements imposed on the physically implementable maps.

The relocation of the error mitigation module to the post-processing stage is possible thanks to informationally complete (IC) measurements at the output of the quantum processing unit. The outcomes of IC measurements can be readily converted into estimates of different observables even if the number of outcomes is much less than what is needed for the state tomography. This is aligned with the approximate reconstruction of a quantum state by using its classical shadows \cite{huang-2020,hadfield-2022,huang-2021}. Not only local but also non-local observables can be estimated this way if IC measurements are optimised~\cite{garcia-perez-2021}. In this regard, the term ``informational completeness'' should not be confused with the number of statistical samples, i.e., the number of measurement outcomes does not generally grow exponentially in the number of qubits. 

The noise inversion map is not completely positive and therefore cannot be directly implemented in quantum hardware (hence, the quasiprobability interpretation has been previously utilised in PEC) but this map can be implemented \emph{in silico} at the classical post-processing stage. This enables us to use the full functionality of tensor network methods that are known to be scalable and well developed for classical simulations of quantum systems~\cite{perez-2007,verstraete-2008,schollwock-2011,hubig-2017,montangero-2018,cirac-2021,evenbly-2022}. The complexity of the constructed noise-mitigation tensor network is reflected in its bond dimension. We show that the lower the noise in the device, the smaller the bond dimension sufficient to precisely mitigate the errors. Truncating the least contributing bonds in the canonical form of a tensor network makes it possible to mitigate the most relevant noise components and maintain a reasonable level of computational complexity. 

The aim of this paper is to provide a full description of the scalable tensor-network error mitigation (TEM) algorithm capable of mitigating quantum noise during the post-processing stage. 

\begin{figure*}
    \centering
    \includegraphics[width = 18cm]{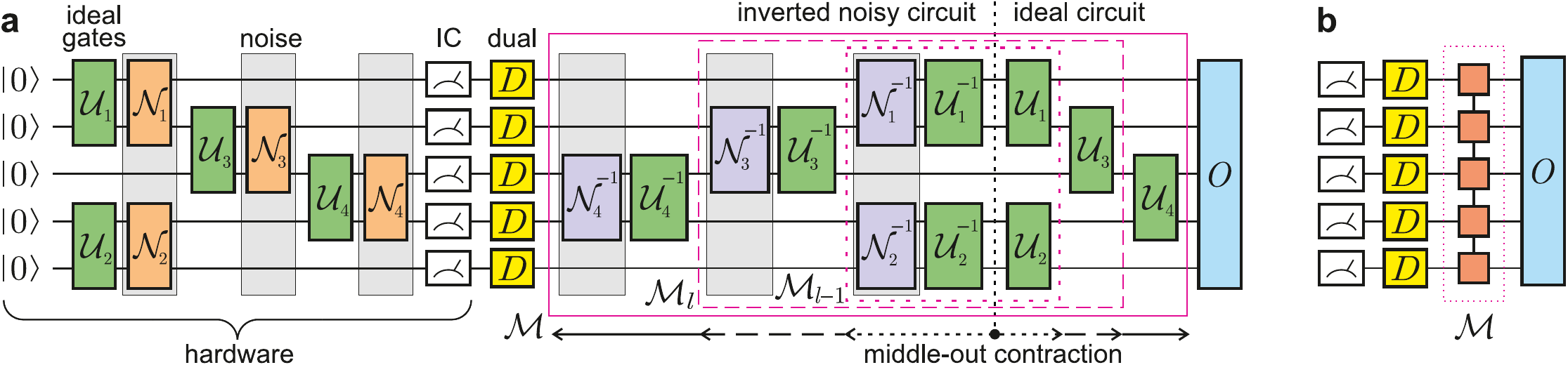}
    \caption{(a) Error-mitigated estimation of an observable $O$ via post-processing measurement outcomes of the noisy quantum processor. ${\cal U}$ and ${\cal N}$ denote an ideal quantum operation and the associated noise map, which can be generally non-local (and extended to grey boxes). $D$ stands for a tensor of operators that are dual to the effects in the IC measurement. The noise mitigation module ${\cal M}$ is a tensor network that is efficiently contracted from the middle out.  The first iteration of the contraction is represented by the dotted purple line, the second one by the dashed line, the third one by the solid line. (b) Tensor network contraction yields the noise-mitigated estimation $\bar{O}_{\rm n.m.}$. Omitting ${\cal M}$, one gets a noisy estimation $\bar{O}_{\rm noisy}$.}
    \label{figure-circuit}
\end{figure*}

As all other noise mitigation strategies, TEM has a side effect in the form of the measurement overhead (needed to keep precision while estimating a desired observable). The advantage of the proposed TEM is that the associated measurement overhead is less than that for PEC. For typical observables we get a quadratic advantage in the measurement overhead.

\section{Noisy quantum computation and IC measurements}

A typical quantum simulator is based on the circuit implementation of quantum computation, where the qubits are initialised in the pure state $\ket{0}^{\otimes N}$, then subjected to unitary gates $U$ from some set of available gates, and finally measured (see Fig.~\ref{figure-circuit}). The set of gates reflects the hardware restrictions such as connectivity of qubits and availability of arbitrary single-qubit unitaries. The purpose of the hardware quantum simulator is to prepare a (highly) entangled quantum state which would be difficult to simulate otherwise (by using a classical computer).

To estimate the average value of some physical observable $O$, the hardware-prepared state $\varrho$ is to be measured. The outcomes of quantum measurements are known to have a probabilistic nature, and the corresponding mathematical description is given by the positive operator-valued measure (POVM). We consider an IC POVM, whose effects span the whole space of operators acting on all $N$ qubits available in the quantum processor. This can be achieved by using IC POVM for each individual qubit, for instance through randomly chosen local projective measurements (Appendix~\ref{appendix-POVM}). 

If $O$ has components with a low Pauli weight (i.e., the Pauli string operators primarily contain identity operators), then there is no need to collect exponentially many (in $N$) measurement outcomes~\cite{garcia-perez-2021}. The number of measurement shots $S$ necessary to estimate the average value ${\rm tr}[\varrho O]$ with precision $\delta$ scales polynomially in $N$ and linearly in $\delta^{-2}$. 

For a finite set of $S$ measurement shots, the estimate $\bar{O}$ for ${\rm tr}[\varrho O]$ and its standard deviation $\Delta \bar{O}$ are given in Appendix~\ref{appendix-estimation}. The idea is to associate each measurement shot ${\bf k}$ with an operator $D_{\bf k}$ dual to the corresponding POVM effect (see Appendix \ref{appendix-POVM}) so that $\bar{O} = \frac{1}{S}\sum_{\bf k} {\rm tr}[D_{\bf k} O]$. In the case of local measurements, each measurement shot ${\bf k} = (k_0,\ldots,k_{N-1})$ is composed of outcomes for individual qubits and $D_{\bf k} = \bigotimes_{m=0}^{N-1} D_{k_m}^{[m]}$ (see Fig.~\ref{figure-circuit}, where local dual operators $\{D_{k_m}^{[m]}\}_{k_m}$ for the $m$th qubit form a tensor $D$). Post-processing of the measurement outcomes and calculation of the estimates $\bar{O}$ and $\Delta \bar{O}$ become particularly straightforward if all the operators are in the Pauli transfer matrix (PTM) representation (Appendix \ref{appendix-PTM}) and the tensor network contractions are utilised (Appendix \ref{appendix-tensor-networks}). Appending the dual tensors to the measurement outcomes, we get an unbiased estimation $\frac{1}{S}\sum_{\bf k} D_{\bf k}$ for the noisy density operator $\varrho$ at the noisy circuit output; the corresponding tensor network is presented in Appendix \ref{appendix-rho}. 

If the observable $O$ is a single Pauli string, it has a trivial tensor network description with disconnected tensors. Otherwise, $O$ can be defined as a sum of weighted Pauli strings, with the corresponding tensor networks being presented in Appendix \ref{appendix-H}. Without any noise mitigation, calculation of the estimate $\bar{O}_{\rm noisy}$ and its standard deviation $\Delta \bar{O}_{\rm noisy}$ reduces to a simple tensor network contraction (Appendix \ref{appendix-tn-contraction}).

In noisy circuits, none of the preparation, dynamics, and measurement steps may be perfect, but the preparation and measurements errors can be relocated to the dynamics part, where the most errors emerge. For this reason we will assume that the initialisation and the measurements are perfect so that all the noise is attributed to the gates. The noisy gate is described by a concatenation ${\cal N} \circ {\cal U}$ of the perfect unitary transformation ${\cal U} [\bullet] = U \bullet U^{\dagger}$ and the completely positive and trace preserving map ${\cal N}$ that can either act on the same qubits as ${\cal U}$ does or affect more qubits due to the decoherence of idle qubits and the interqubit crosstalk. Our noise mitigation method allows the noise ${\cal N}$ to affect all qubits in the register provided the detailed and compact description of this map is given, e.g., in the form of the one-dimensional tensor network with topology of the locally-purified density operator~\cite{werner-2016,torlai-2023,srinivasan-2021} also known as the matrix product channel~\cite{mpc-2022}. The key requirement for the proposed TEM algorithm is that the inverse map ${\cal N}^{-1}$ has a compact tensor network representation with a modest bond dimension. This requirement is naturally met in a number of practical scenarios discussed in Sec.~\ref{section-middle-out}.

\section{Tensor-network error mitigation} \label{section-middle-out}

A noisy quantum circuit with the characterised noise is measured with an IC POVM, and the collected measurement outcomes are processed on a classical computer. This enables us to formally perform all mathematical transformations (including a non-physical map ${\cal N}^{-1}$) and employ the tensor-network methods. Fig.~\ref{figure-circuit} depicts the proposed TEM map ${\cal M}$ that fully inverts the whole noisy circuit and applies the ideal noiseless circuit, i.e. the noisy circuit output $\varrho$ is reverted back to $(\ket{0}\bra{0})^{\otimes N}$, which in turn is mapped to the noiseless operator $\ket{\psi}\bra{\psi}$. The noise-mitigated estimation of an observable $O$ reads $\bar{O}_{\rm n.m.} = \frac{1}{S}\sum_{\bf k} {\rm tr}[{\cal M}(D_{\bf k}) O] = \frac{1}{S}\sum_{\bf k} {\rm tr}[D_{\bf k} {\cal M}^{\dag}(O)]$. Once we have a compact tensor network description for ${\cal M}$, calculation of $\bar{O}_{\rm n.m.}$ and $\Delta\bar{O}_{\rm n.m.}$ again reduces to a simple tensor network contraction (Appendix \ref{appendix-tn-contraction}).

If one na\"{\i}vely builds the noise mitigation map ${\cal M}$ by concatenating all the maps layer by layer [i.e., ${\cal M} = (\bigcirc_{l} {\cal U}_l) \circ \bigcirc_{l} ({\cal U}_l^{-1} \circ {\cal N}_l^{-1} )$], then this is even more demanding than simulating a noiseless quantum computation on a classical computer. However, the calculation of ${\cal M}$ can be made computationally efficient and sufficiently accurate by exploiting the fact that every unitary layer ${\cal U}_l$ and the corresponding map ${\cal U}_l^{-1} \circ {\cal N}_l^{-1}$ approximately cancel each other ensuring a tensor network representation with a low bond dimension. 

\begin{figure}
    \centering
    \includegraphics[width = 8.5cm]{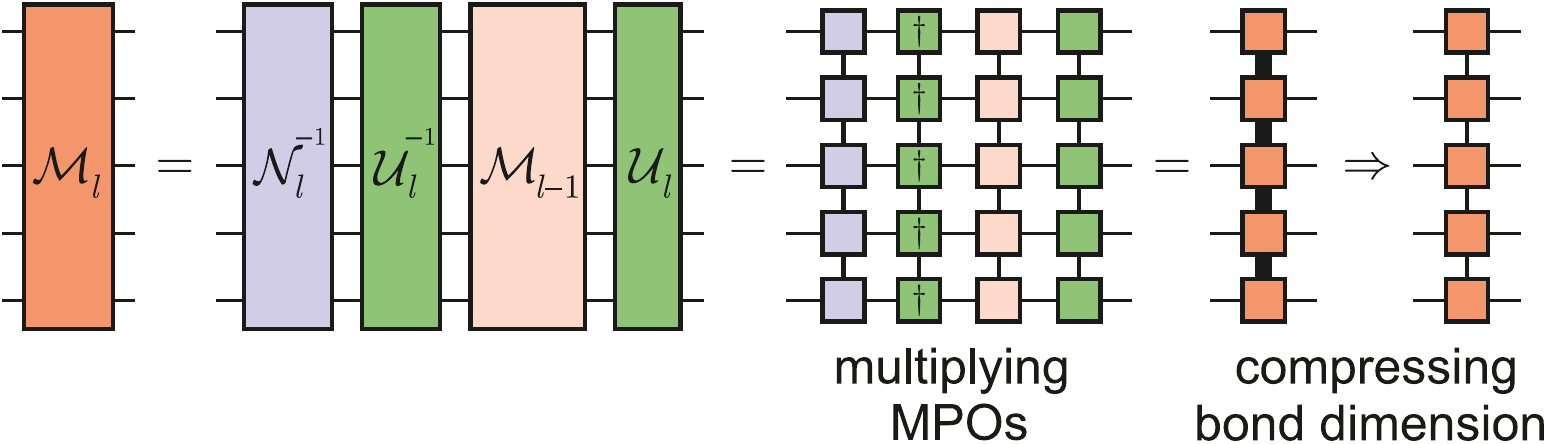}
    \caption{One iteration in the middle-out contraction: multiplication of layer MPOs and the bond dimension truncation (MPO compression).}
    \label{figure-mpos}
\end{figure}

So the map ${\cal M}$ is considered as a tensor network, whose contraction starts from the middle (where the inverted noisy circuit ends and the ideal circuit starts) and propagates outwards by involving two layers on the left side and one layer on the right side at each iteration. A single iteration reads
\begin{equation} \label{middle-out-iteration}
    {\cal M}_{l} = {\cal U}_l \circ {\cal M}_{l-1} \circ {\cal U}_l^{-1} \circ {\cal N}_l^{-1}. 
\end{equation}
As we explain below, all the maps ${\cal U}_l$, ${\cal U}_l^{-1}$, and ${\cal N}_l^{-1}$ adopt a computationally-efficient form of the matrix-product-operator (MPO) tensor network of linear topology depicted in Fig.~\ref{figure-mpos}. An MPO of bond dimension $\chi$ for an $N$-qubit map has the form
\begin{equation}
    \sum_{a_0,\ldots,a_{N-2}=0}^{\chi-1} {\cal A}^{[0]}_{a_0} \otimes {\cal A}^{[1]}_{a_0 a_1} \otimes \cdots \otimes {\cal A}^{[N-2]}_{a_{N-3} a_{N-2}} \otimes {\cal A}^{[N-1]}_{a_{N-2}}, 
\end{equation}
where, for any fixed values of virtual indices $a_{m-1}$ and $a_m$, ${\cal A}^{[m]}_{a_{m-1} a_m}$ is a map acting on the $m$th qubit. Each iteration in Eq.~\eqref{middle-out-iteration} reduces to a standard multiplication of MPOs that yields an MPO with a multiplicative bond dimension (Appendix \ref{appendix-multiplication-mpos}).
The memory cost of storing an $N$-qubit MPO scales as $\chi_{\rm max}^2 N$ and the computational cost of MPO multiplication is of the same order due to its triviality.

Consider a typical hardware circuit composed of single-unitary gates and layers of non-overlapping {\sc cnot} gates in a linear-connectivity layout. Then the MPO form for a unitary layer ${\cal U}_l$ has the bond dimension $4$ and immediately follows from a trivial decomposition of each {\sc cnot} superoperator (Appendix \ref{appendix-mpo-unitary}). Needless to say, the MPO for ${\cal U}_l^{-1}$ is readily obtained from the MPO for ${\cal U}_l$ via conjugation and has the same bond dimension. 

The noise-inversion map ${\cal N}_l^{-1}$ is efficiently represented by an MPO with a modest bond dimension either from a local structure of the known noise (tomography of individual noisy gates), or through a characterised Pauli-Lindblad model with crosstalk \cite{berg-2023} (describing the noise after applying the randomised compiling technique \cite{geller-2013,wallman-2016}), or via inversion \cite{guo-2022} of the matrix-product-channel noise inferred by the method of Ref.~\cite{torlai-2023}. In the case of the Pauli-Lindblad model with a nearest-neighbour crosstalk, the noisy layer ${\cal N}_l$ is efficiently represented as a sequence of commuting two-qubit Pauli channels applied to adjacent qubits, with the resulting MPO for ${\cal N}_l^{-1}$ having bond dimension $\chi_{\rm n} = 4$ (Appendix \ref{appendix-noise-mpo}). Should the two-qubit channels be depolarising, then the MPO bond dimension for ${\cal N}_l^{-1}$ reduces to $\chi_{\rm n} = 2$ (Appendix \ref{appendix-noise-mpo}). 

Assuming ${\cal N}_l^{-1}$ has some bond dimension $\chi_{\rm n}$ and the current iteration ${\cal M}_{l-1}$ has bond dimension $\chi_{l-1}$, the next iteration map ${\cal M}_l$ in Eq.~\eqref{middle-out-iteration} has bond dimension $\chi_l = 16 \chi_{\rm n} \chi_{l-1}$. This results in an exponentially growing bond dimension $16^L \chi_{\rm n}^{L-1}$ for a circuit of depth $L$. To overcome this difficulty, the MPO is compressed after each iteration either to a fixed bond dimension at most $\chi_{\rm max}$ or to a desired precision. This is achieved by truncating the smallest singular values in the canonical representation for the MPO or by variational methods \cite{hubig-2017}.

The most crucial feature of the proposed from-the-middle-out contraction for ${\cal M}$ is that it captures cancellation effects for unitaries and their inverses when the noise level is reasonably small (i.e., the map ${\cal N}$ is close to the identity transformation ${\rm Id}$). Then ${\rm Id}$ is the leading contribution in ${\cal M}$, and ${\rm Id}$ is known to have a trivial bond dimension $1$. When the noise level $\epsilon$ is small but non-zero, the expansion ${\cal N} \approx {\rm Id} + \epsilon \Lambda$ leads to ${\cal U} \circ {\cal N}^{-1} \circ {\cal U}^{-1} \approx {\rm Id} - \epsilon {\cal U} \circ \Lambda \circ {\cal U}^{-1}$. Continuing this line of reasoning for iterative Eq.~\eqref{middle-out-iteration}, we see that the second largest singular value in every MPO link for $\cal M$ is of the order of $\epsilon$. As a result, the MPO compression error is at most linear in $\epsilon$. Numerical analysis of the singular values of MPO links justifies this observation (Appendix \ref{appendix-sv-distribution}). For the sufficiently large bond dimension $\chi_{\rm max}$ exceeding a certain threshold, the compressed MPO reproduces all first-order singular values of the exact MPO, and then the truncation error exhibits a transition to the order of $\epsilon^2$. The computational cost of the MPO compression scales as $\chi_{\rm max}^3 N$ \cite{hubig-2017}, and it thus outweighs the cost of the MPO multiplication. 

\section{Measurement overhead in TEM}

Similarly to PEC, TEM amplifies the measurement shot noise. The measurement overhead $\gamma$ is the ratio of standard deviations in estimations of the observable after and prior to the noise mitigation. $\gamma^2$ is the scaling factor in the number of shots needed to get a desired estimation precision and quantifies quantum computational resources. In PEC, the measurement overhead originates from the physical simulation of ${\cal N}^{-1}$ via sampling and averaging over unitary operations from the quasiprobability representation for ${\cal N}^{-1}$. Suppose the noise is a mixture ${\cal N} = (1-\epsilon) {\rm Id} + \epsilon \Lambda_{\rm P}$, where $\Lambda_{\rm P}[\bullet] = \sum_{\boldsymbol{\alpha} \neq {\bf 0}} p_{\boldsymbol{\alpha}} \sigma_{\boldsymbol{\alpha}} \bullet \sigma_{\boldsymbol{\alpha}}$ is a random unitary quantum channel and unitary operators are non-trivial Pauli strings $\sigma_{\boldsymbol{\alpha}} = \bigotimes_{m=0}^{N-1} \sigma_{\alpha_m}$. Then ${\cal N}^{-1} \approx (1+\epsilon) {\rm Id} - \epsilon \Lambda_{\rm P}$ and $\gamma_{\rm PEC} \approx 1 + 2 \epsilon$ for this layer. PEC has two summands in the measurement overhead: the amplifying factor $1+\epsilon$ and the negativity-induced contribution $\epsilon$.

In TEM, ${\cal N}^{-1}$ is a purely mathematical map kept in the memory of a classical computer, so the absence of complete positivity does not cause a problem. The dual map $({\cal N}^{-1})^{\dag}$ formally describes the evolution of an observable $O$ in the Heisenberg picture [in the case above we have self duality, $({\cal N}^{-1})^{\dag} = {\cal N}^{-1}$]. For a Pauli string observable $O = \sigma_{\boldsymbol{\beta}}$, we have $({\cal N}^{-1})^{\dag}[\sigma_{\boldsymbol{\beta}}] = \gamma_{\rm TEM} \sigma_{\boldsymbol{\beta}}$ with the measurement overhead $\gamma_{\rm TEM} \approx 1+\epsilon - \epsilon \sum_{\boldsymbol{\alpha} \neq {\bf 0}} p_{\boldsymbol{\alpha}} (-1)^{\langle \boldsymbol{\alpha}, \boldsymbol{\beta} \rangle}$, where $\langle \boldsymbol{\alpha}, \boldsymbol{\beta} \rangle = 0~(1)$ if $\sigma_{\boldsymbol{\alpha}}$ and $\sigma_{\boldsymbol{\beta}}$ commute (anticommute). We see that $\gamma_{\rm TEM} \leq \gamma_{\rm PEC}$ always holds, and $\gamma_{\rm TEM} < \gamma_{\rm PEC}$ whenever $\sigma_{\boldsymbol{\beta}}$ commutes with at least one of the $\sigma_{\boldsymbol{\alpha}}$ for which $p_{\boldsymbol{\alpha}} \neq 0$. If $\sigma_{\boldsymbol{\beta}}$ commutes with all of the non-trivially contributing operators $\sigma_{\boldsymbol{\alpha}}$ (e.g., $\sigma_{\boldsymbol{\beta}}=I$), then $\gamma_{\rm TEM} = 1$. In experimental studies of the Pauli-Lindblad noise models \cite{berg-2023}, the distribution $\{p_{\boldsymbol{\alpha}}\}_{\boldsymbol{\alpha}}$ is roughly flat, which means that typically commuting and anticommuting terms cancel each other ($\sum_{\boldsymbol{\alpha} \neq {\bf 0}} p_{\boldsymbol{\alpha}} (-1)^{\langle \boldsymbol{\alpha}, \boldsymbol{\beta} \rangle} \approx 0$) and $\gamma_{\rm TEM} \approx 1 + \epsilon \approx \sqrt{\gamma_{\rm PEC}}$ if $\boldsymbol{\beta} \neq {\bf 0}$. Dealing with a general observable $O = \sum_{\boldsymbol{\beta}} c_{\boldsymbol{\beta}} \sigma_{\boldsymbol{\beta}}$, where the expansion coefficients $\{c_{\boldsymbol{\beta}}\}$ are all of the same order, we again get the averaged overhead $\gamma_{\rm TEM} \approx 1 + \epsilon \approx \sqrt{\gamma_{\rm PEC}}$. Explicit examples for the quadratic improvement in the measurement overhead are given in Appendix \ref{appendix-measurement-overhead}. Since the measurement overhead scales exponentially with the circuit depth $L$, the improvement in the measurement overhead becomes drastic for deep circuits: $\gamma_{\rm TEM} \approx (1 + \epsilon)^L \ll (1 + 2\epsilon)^L \approx \gamma_{\rm PEC}$.

An observable with a low Pauli weight (the components of which act trivially on all but a few qubits) has a smaller measurement overhead both in PEC \cite{tran-2023} and TEM due to the causal cone structure in its Heisenberg evolution and the commutation with many low-weight Pauli strings in $\Lambda_{\rm P}$ (provided the crosstalk noise affects nearest qubits). However, the relation $\gamma_{\rm TEM} \approx \sqrt{\gamma_{\rm PEC}}$ still remains because TEM has one contribution in the measurement overhead in contrast to two contributions in PEC.

\section{Results} \label{section-results}

\begin{figure*}
    \centering
    \includegraphics[width = 18cm]{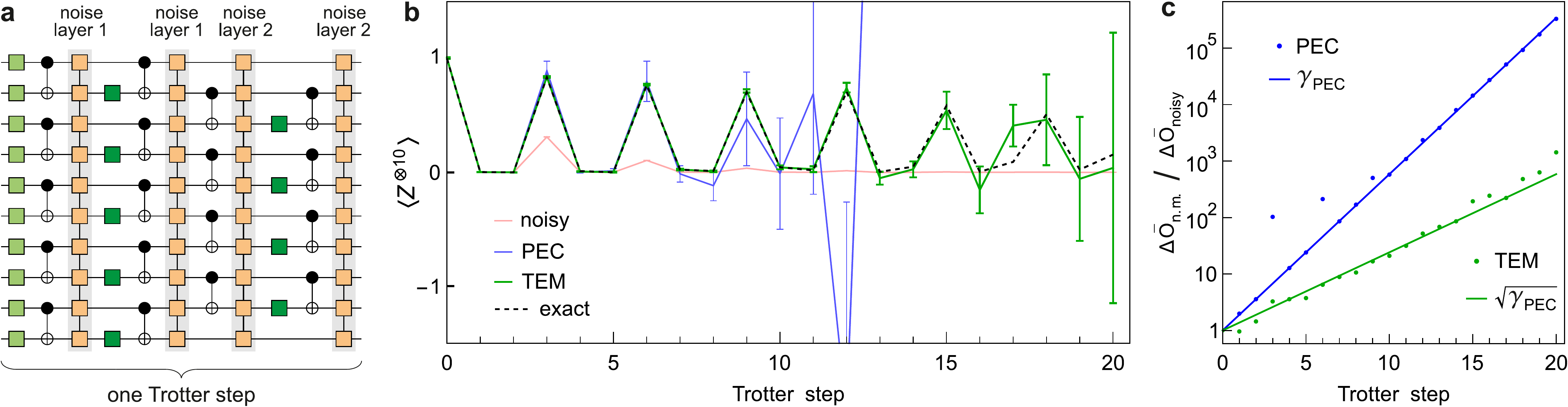}
    \caption{Comparison of TEM and PEC in estimating the non-local observable $O=Z^{\otimes 10}$ in the 10-qubit discrete-time evolution. (a) A single Trotter step for the one-dimensional transverse-field Ising model contains a unitary rotation $R_X(2h\delta_t)$ for every qubit and two distinct implementations of pairwise $ZZ$-rotations (for even and odd links between the qubits), each consisting of two repeated {\sc cnot} layers intervened by the unitary rotation $R_Z(-2J\delta_t)$ on controlled qubits. Model parameters $h = 1$, $J = 0.5236$, and $\delta_t = 0.5$ reproduce the setting of Ref.~\cite{berg-2023}. Each unique {\sc cnot} layer is followed by a sparse Pauli-Lindblad noise (MPO of bond dimension 4) with $\gamma_{\rm PEC} = 1.183$ for the noise layer 1 and $\gamma_{\rm PEC} = 1.162$ for the noise layer 2. (b) Dynamics of the parity observable estimation without and with noise mitigation. PEC estimation is based on 300 circuits sampled from the quasiprobability representation of the inverse noise and 10000 shots per circuit in the computational basis. TEM estimation is based on 300 circuits with projective measurements in either $X$-, or $Y$-, or $Z$-basis for every qubit and 10000 shots per circuit. The measurement bases are chosen with the probabilities $p_x=0.001$, $p_y=0.001$, $p_z=0.998$ (see IC POVM description in Appendix~\ref{appendix-POVM}) to adjust for the heavy-weight observable of interest. Error bars correspond to the estimation errors $\Delta \bar{O}_{\rm n.m.}$, computed using Eq.~\eqref{error-formula-settings}. Bond dimension of the noise mitigation map is at most 400. (c) Measurement overhead in the numerical experiment (dots) as the ratio of the noise-mitigated and unmitigated estimation errors, $\Delta \bar{O}_{\rm n.m.} / \Delta \bar{O}_{\rm noisy}$, and the theoretical prediction (lines).}
    \label{figure-trotter}
\end{figure*}

First, we show TEM for the discrete-time 10-qubit dynamics recently simulated on a digital quantum computer with PEC \cite{berg-2023}. A single Trotter step for the one-dimensional transverse-field Ising model is depicted in Fig.~\ref{figure-trotter}a. We assume that the single-qubit gates are perfect and the noise in each of two unique {\sc cnot} layers is characterised and given in the form of an individual sparse Pauli-Lindblad model \cite{berg-2023}. The model parameters (111 decoherence rates in Appendix~\ref{appendix-rates}) are chosen in such a way that the measurement overhead is comparable with the experiment in Ref.~\cite{berg-2023}. With this noise level and no error mitigation, the noisy estimation of the parity observable $O=Z^{\otimes 10}$ via standard measurements in the computational basis quickly decays and significantly deviates from the exact value after Trotter step 2, see Fig.~\ref{figure-trotter}b. Both PEC and TEM rectify the estimation but require different computational resources. 

To compare the two methods on an equal footing, we reserve the same number of measurement shots. In PEC simulation, we assume the noise is already twirled to the sparse Pauli-Lindblad form and sample only unitary gates from the quasiprobability distribution of the inverse noise maps. In TEM simulation, the IC POVM is implemented through sampling projective measurements in different bases (Appendix \ref{appendix-POVM}), so we use the same number of circuits and shots per circuit as in PEC (see these values in the caption of Fig.~\ref{figure-trotter}). In this setting, PEC and TEM adequately reproduce the exact values of the observable until the measurement overhead spoils the standard deviation $\Delta \bar{O}_{\rm n.m.}$ in the mitigated observable. Remarkably, TEM provides accurate estimations for twice as many Trotter steps as compared to PEC (for a fixed value of $\Delta \bar{O}_{\rm n.m.}$). This is due to the quadratic improvement in the measurement overhead inherent in TEM. We visualise the ratio of the noise-mitigated and unmitigated estimation errors, $\Delta \bar{O}_{\rm n.m.} / \Delta \bar{O}_{\rm noisy}$, in Fig.~\ref{figure-trotter}c. The simulation results agree with the theoretical predictions for the measurement overhead and confirm the relation $\gamma_{\rm TEM} \approx \sqrt{\gamma_{\rm PEC}}$.

\begin{figure*}
    \centering
    \includegraphics[width = 18cm]{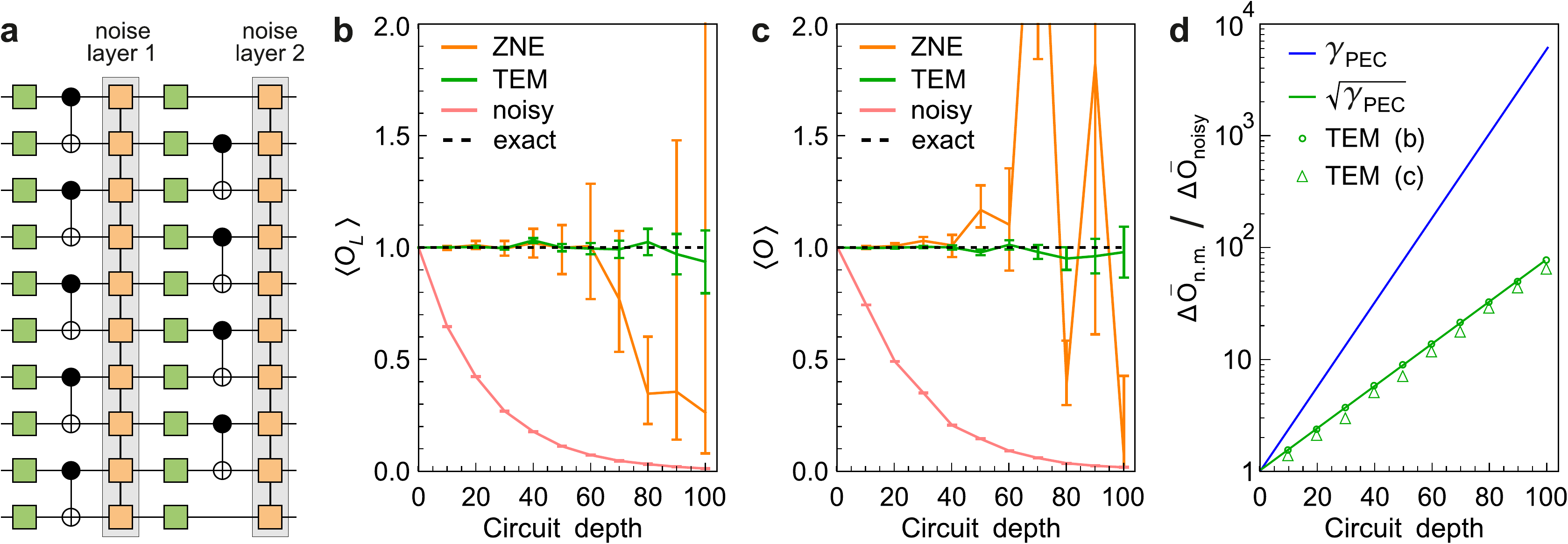}
    \caption{Comparison of some noise-mitigation strategies. (a) Two layers of the brickwork stabiliser circuit. The considered circuit has 100 qubits. Each unique {\sc cnot} layer is followed by a sparse Pauli-Lindblad noise (MPO of bond dimension 4) with the effective noise per layer per qubit $\bar{\gamma}_{\rm PEC}-1 = 9.1 \times 10^{-4}$. (b,c) Noisy and noise-mitigated estimations of the stabiliser operator for the output quantum state. The operator has a high relative Pauli weight in (b), e.g., $O_L=X_{2,3,6,17,18,23,31,37,39,41,45,46,47,51,56,61,64,65,67,69,73,84,90,99}$$Y_{4,7,9,10,12,14,15,16,29,30,33,36,38,43,53,55,58,60,63,68,70,72,76,77,79,81,85,87,91,93,96}$ $Z_{5,21,22,32,34,40,44,49,52,54,59,66,88,89,94,97,100}$ for some circuit depth $L$, and a low relative Pauli weight in (c), $O=X_{10,20,28,29,38,48,49,50,57,69,70,71,77,90,91,92,96}Y_{11,30}Z_{100}$. In ZNE, 3 extrapolation points with $3 \times 10^5$ shots per point are used. TEM is run with the the bond dimension $700$ and $3 \times 10^5$ shots. (d) Measurement overhead in PEC and TEM for both examples.}
    \label{figure-stabilizer-results}
\end{figure*}

As the second example we consider stabiliser quantum circuits that serve as a natural testbed for studying the scalability of quantum informational protocols. Fig.~\ref{figure-stabilizer-results}a depicts 2 layers of the stabiliser circuit with the brickwork {\sc cnot} gates and randomly chosen single-qubit Clifford gates. The circuit consists of $L$ brickwork layers and defines a unitary operator $U_L$. At the circuit output, the observable $O_L := U_L G U_L^{\dag}$ has expectation value $1$ for any stabiliser $G$ of the initial state ($G\ket{0}^{\otimes N} = \ket{0}^{\otimes N}$). 

We simulate noise in each {\sc cnot} layer by a sparse Pauli-Lindblad model with the measurement overhead $\bar{\gamma}_{\rm PEC}^{N}$, so that $\bar{\gamma}_{\rm PEC}-1$ quantifies the effective noise per layer per qubit. For instance, if each {\sc cnot} is followed by 2-qubit depolarising noise of intensity $\epsilon \ll 1$, then $\bar{\gamma}_{\rm PEC} \approx 1 + \tfrac{15}{16}\epsilon$ (Appendix \ref{appendix-pauli-lindblad-noise-model}). We perform projective measurements in the eigenbasis of each Pauli operator in the string $O_L$ because the eigenstates of ${\cal N}(\sigma_{\boldsymbol{\alpha}})$ and $\sigma_{\boldsymbol{\alpha}}$ coincide for the considered noise ${\cal N}$. This can be viewed as an optimisation of IC-POVMs where non-signalling effects have vanishing weights (see Appendix \ref{appendix-stabilizer} for more details).

For deep circuits, $O_L$ has a high Pauli weight whatever initial stabiliser $G$ is chosen. We choose $G=Z^{\otimes N}$ so that $O_L$ has high Pauli weight for all depths $L$. Fig.~\ref{figure-stabilizer-results}b depicts the results the TEM and ZNE for 100 qubits and different depths $L$, up to $L= 100$ (for details on the numerical simulations, see Appendix~\ref{app:numerical_details}). With the same budget of shots for the TEM and each noise gain parameter in ZNE, the latter exhibits extrapolation instability for $L \geq 60$ because the noisy and noise-amplified observable estimates cluster around $0$ (Appendix \ref{appendix-zne}). In turn, PEC cannot benefit from causal cones as they are absent for the considered observable \cite{tran-2023}, so the total measurement overhead $\gamma_{\rm PEC} = \bar{\gamma}_{\rm PEC}^{NL}$ grows exponentially in $L$. Fig.~\ref{figure-stabilizer-results}d shows that the measurement overhead in TEM also grows exponentially but with a halved power, $\gamma_{\rm TEM} = \bar{\gamma}_{\rm PEC}^{NL/2}$, manifesting the quadratic advantage over PEC and a prominent improvement especially for deep circuits.

Keeping a quantum chemical Hamiltonian in mind, we also consider a low-weight Pauli string as an observable $O$ (representing a single-excitation term in the Hamiltonian using the optimal fermion-to-qubit mapping from Ref.~\cite{jiang-2020}). To get a non-zero estimate for such an observable in the considered Clifford circuit, we propagate the observable in the Heisenberg picture ($G_L := U_L^{\dag} O U_L$) and locally modify the initial state $\ket{0}^{\otimes N}$ so that it is stabilised by $G_L$. Thanks to a lower Pauli weight of $O$, its noisy estimate is improved (as compared to $O_L$) but not significantly due to the scattered nature of non-trivial Pauli operators in $O$ (vertices for the causal cones), see Fig. \ref{figure-stabilizer-results}c. The measurement overhead in this case is slightly less than in the case of the heavy-Pauli-weight observable but is aligned with $\sqrt{\gamma_{\rm PEC}}$: this is the effect of causal cones that start overlapping at circuit depth $\sim 10$.

\section{Discussion} \label{section-discussion}

\begin{figure}
    \!\!\!\!\!\!\!\!\!\!\!\!\includegraphics[width=10cm]{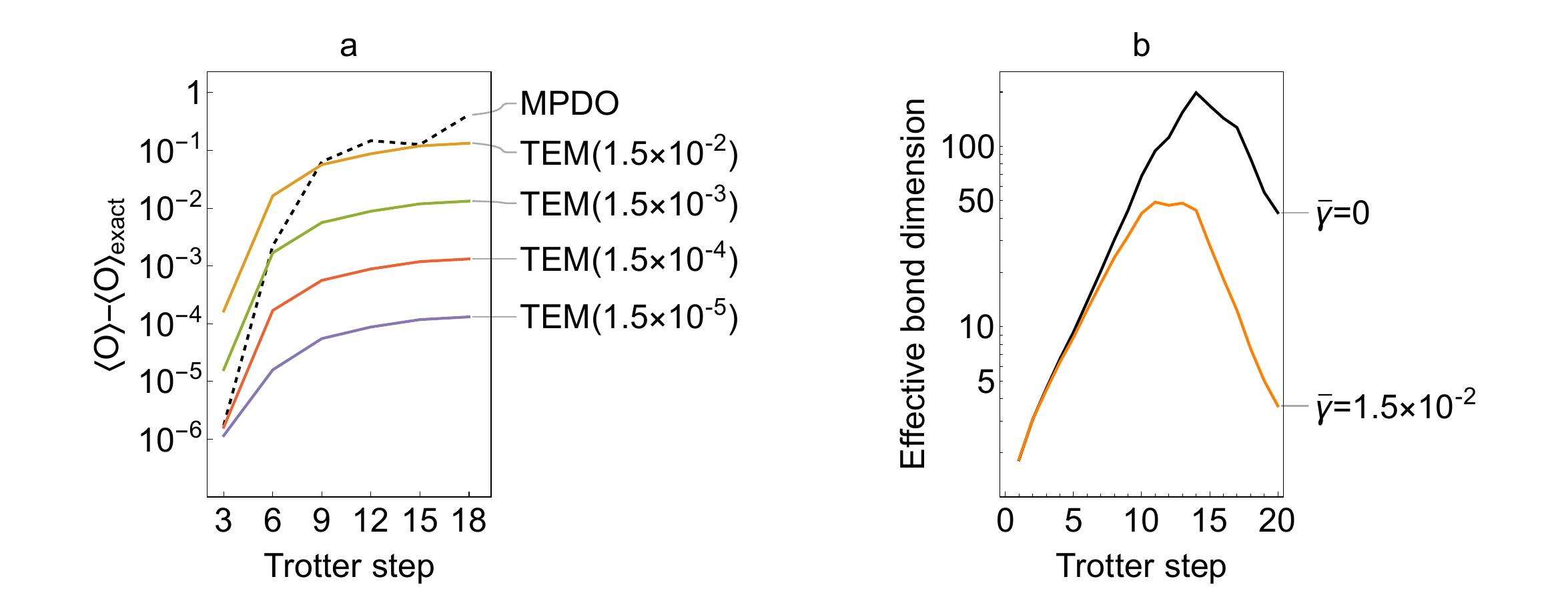}
    \caption{TEM and classical simulation of quantum dynamics presented in Fig.~\ref{figure-trotter}. (a) The effect of the truncation error in TEM module with bond dimension 100 on the observable estimation error, $\bar{O}_{\rm n.m.}(S=\infty) - \braket{O}_{\rm exact}$, for different levels of noise ($\bar{\gamma}_{\rm PEC}$ per layer per qubit). Classical estimation with MPDO of bond dimension 100. (b) The effective bond dimension $(\sum_i \lambda_i^2)^{-1}$ expressed through normalised singular values in the central MPDO link ($\sum_i \lambda_i = 1$) for exact noiseless and noisy dynamics.}
    \label{figure-error-bds}
\end{figure}

We have demonstrated that the tensor network methods can be used to mitigate errors in noisy quantum processors. On the other hand, tensor network methods are used in purely classical simulations of quantum circuits too. A natural question arises whether TEM is able to outperform a classical tensor network simulation provided their bond dimensions coincide. In Fig. \ref{figure-error-bds} we provide an affirmative answer to this question by considering a toy 10 qubit example with a modest bond dimension for both TEM and the matrix-product-density-operator (MPDO) classical simulation. Provided the shot noise is suppressed to the extent that only the truncation error affects the observable estimation, TEM provides a more accurate estimate than the classical simulation for noise levels below those in the experiment \cite{berg-2023}. The less the noise in a quantum processor, the preciser TEM estimation. In the presented example, the truncation error is linear in the noise strength as expected for a rather small bond dimension. TEM needs to amend only those correlations that were effectively destroyed by noise in the device. Computational complexity is distributed between the noisy quantum processor and the TEM module, so the more correlations survive in the quantum processor, the less bond dimension is needed in TEM to achieve a desired precision. 

As we have shown, applying the noise mitigation map virtually in post-processing, as opposed to decomposing it in terms of physical operations to be performed on hardware, requires significantly less quantum resources.
The price to pay is the classical computational cost associated with the middle-out compression of the map, embodied in the bond dimension $\chi$ needed to get accurate mitigated results. 
Unfortunately, estimating the required bond dimension {\it a priori} is not an easy task.
While a polynomially scaling bond dimension is expected to suffice for accounting for the most significant terms in the map, as it stems from a perturbation theory analysis of the method (Sec.~\ref{section-middle-out}), the specific value of $\chi$ depends not only on the noise $\epsilon$, but also on the observable and the number of shots used.

On the one hand, notice that the bond dimension needed to capture a specific observable well enough may be significantly lower than the bond dimension needed to represent the whole map $\mathcal{M}_L$ accurately.
The reason is that $\mathcal{M}_L$ is a superoperator that can be used to compute ${\rm tr}[\mathcal{M}_L(\varrho)O]$ for any input state $\varrho$ and observable $O$.
In TEM, we truncate the map to prevent an exponential scaling of the bond dimension, so we use a map $\mathcal{M}'_L$ that is similar but not equal to $\mathcal{M}_L$.
As a consequence ${\rm tr}[\mathcal{M}'_L(\varrho)O]$ will be close but not equal to ${\rm tr}[\mathcal{M}_L(\varrho)O]$, and the error $\Delta(\varrho, O) = \vert {\rm tr}[\mathcal{M}'_L(\varrho)O] - {\rm tr}[\mathcal{M}_L(\varrho)O] \vert$ may be different for different $\varrho$ and $O$.
Yet, we are generally not interested in applying the map for any other $\varrho$ than the noisy state of the quantum processor.
Similarly, we are generally only interested in a subset of all possible observables $O$.
Therefore, we do not need to approximate the full map $\mathcal{M}_L$ well enough, but only some relevant sectors of it, which might be a much easier task that demands a smaller $\chi$.

On the other hand, since working with finite statistics is unavoidable, there will always be a statistical error associated with every observable expectation value, even for the exact map $\mathcal{M}_L$.
Therefore, if the estimate of ${\rm tr}[\mathcal{M}_L(\varrho)O]$ has an associated standard error $\Delta \bar{O}$, it suffices that the truncation-related error $\Delta(\varrho, O)$ be smaller than $\Delta \bar{O}$, as higher accuracy would not even be observable without additional statistics.
This fact also simplifies the task of approximating the relevant sector of $\mathcal{M}_L$.

In short, the analysis of the bond dimension needed by TEM in practical applications should not be focused on the analysis of the map $\mathcal{M}_L$ alone, but rather on the specific observable and measurement data obtained.
Crucially, the fact that TEM is unbiased (assuming perfect noise characterisation, as in PEC) for large enough bond dimension can be used to assess the minimal $\chi$ needed in specific situations.
The idea is to monitor the expectation value of the observable for increasing $\chi$ until the changes in the result become smaller than the statistical fluctuations.
In Appendix \ref{appendix-convergence}, we explain the method in detail and provide an example.
Interestingly, we observe that the error in the estimation decreases exponentially with the bond dimension (see Fig.~\ref{figure-convergence}).

The fact that the classical part of TEM is a tensor network method implies that many powerful techniques and advancements from that field can be borrowed for noise mitigation.
For instance, it may be possible to use TEM in situations where, due to high levels of noise, a very large bond dimension may be needed by performing an infinite bond dimension extrapolation, akin to the technique used in Ref.~\cite{tindall-2023}.
Similarly, there is no fundamental reason why TEM must only use MPOs.
Other tensor network structures~\cite{frowis-2010}, such as PEPOs, can prove beneficial, especially in situations where the circuit is not linearly connected.

In addition to improvements on the classical side of TEM, there is also substantial room for combining it with other noise mitigation techniques, such a PEC and ZNE.
For instance, in the case of sparse Pauli-Lindblad noise, PEC could be used only to invert the terms associated with the two-qubit Lindblad operators, which usually have smaller coefficients than the single-qubit ones, while TEM can account for the latter ones. 
In this way, the bond dimension of the noise layer is equal to one, which simplifies the middle-out contraction, while the measurement overhead may not increase substantially. ZNE can benefit from TEM as it can readily mitigate noise partially (with a smaller measurement overhead than PEC), thus enabling the noise scaling parameter $G$ to enter the region in between 0 and 1, where ZNE is much more accurate.

\section{Conclusions} \label{section-conclusions}

Error mitigation is essential to the utility of quantum computing in the near-term, and will be instrumental in the path towards fault-tolerant quantum devices.
The technology is evolving quickly, and the recent demonstrations of the potential of current devices for quantum simulation indicate that useful quantum computing is much closer than expected by a large fraction of the community.
Yet, noise stands as the major challenge to be overcome, and new approaches to error mitigation can have a critical impact in the evolution of the field in the near term.

In this work, we have introduced TEM, a hybrid quantum-classical error mitigation algorithm in which the state of the noisy quantum processor is measured with informationally complete measurements, and the outcomes are then post-processed by a tensor network map that represents the inverse of the global noise channel biasing the results.
TEM presents the crucial feature that the measurement overhead incurred is quadratically smaller than PEC, which allows us to, for instance, double the depth of the circuit that can be accurately mitigated with the latter method.
By contrasting its performance with ZNE in the regime of 100 qubits and depth 100 Clifford circuits with a realistic noise model (sparse Pauli-Lindblad one), we have shown that ZNE suffers from instabilities that make the method unreliable, while TEM succeeds at producing correct expectation values both for high-weight and fermionic-simulation-related Pauli strings.
In turn, in order to reach the levels of accuracy obtained with TEM, PEC would require a prohibitive number of measurement shots.

TEM can also be advantageous with respect to purely classical tensor network methods, given that the tensor network in TEM does not need to account for the state of the quantum computer, nor the evolved observable in the Heisenberg picture.
Instead, the tensor network represents the inverse of the noise channel in the quantum processor, which approaches identity for decreasing noise.
Therefore, the classical computational complexity needed by TEM also decreases, hence enabling to obtain accurate results with smaller computational cost than a classical-only tensor network approach.

In the current context, in which quantum computing and tensor network methods seem to be competing with one another, TEM offers a different, more collaborative perspective, as it will benefit from developments in both fronts.
More importantly, TEM paves the way to a very exciting prospect: the combination of quantum computing and tensor networks can be computationally more powerful than either of the two alone.

\section*{Acknowledgements}
The authors thank Daniel Cavalcanti, Elsi-Mari Borrelli, Marco Cattaneo, Zolt\'{a}n Zimbor\'{a}s, and Sabrina Maniscalco for interesting discussions, and Ivano Tavernelli, Francesco Tacchino, Laurin Fischer, Zlatko Minev, Alireza Seif, Sarah Sheldon, and Christopher Wood for their insights regarding PEC, ZNE, and their realistic implementation on hardware.
The authors are also grateful to Aaron Miller, Keijo Korhonen, Joonas Malmi, and Boris Sokolov for their help with the construction of the chemical Pauli strings, improvements in the code, and pointing out simulator packages for the Clifford circuits.

\section*{Competing interests}
Elements of this work are included in patents filed by Algorithm Ltd with the European Patent Office.

\section*{Author contributions}
SF and GGP conceived the algorithm.
SF and GGP designed and directed the research.
SF, ML, MR and GGP implemented the algorithm and ran simulations.
SF wrote the first version of the manuscript.
All authors contributed to scientific discussions and to the writing of the manuscript.

\bibliography{bibliography}

\clearpage

\appendix
\vspace{2cm}
\onecolumngrid

\renewcommand{\theequation}{S\arabic{equation}}
\renewcommand{\theproposition}{S\arabic{proposition}}
\renewcommand{\thecorollary}{S\arabic{corollary}}
\renewcommand{\thefigure}{S\arabic{figure}}
\setcounter{equation}{0}
\setcounter{table}{0}
\setcounter{section}{0}
\setcounter{proposition}{0}
\setcounter{corollary}{0}
\setcounter{figure}{0}

\numberwithin{equation}{section}

\section{Informationally complete POVM} \label{appendix-POVM}

Informationally complete POVM for a single qubit contains at least $4$ effects $\{\Pi_{k}\}_{k}$. The effects are Hermitian positive semidefinite operators ($\Pi_{k}^{\dag} = \Pi_{k} \geq 0$) summing to the identity operator ($\sum_k \Pi_k = I$). Informational completeness implies that ${\rm dim} \, {\rm Span}(\{\Pi_{k}\}_k) = 4$ and the equality ${\rm tr}[\varrho \Pi_k] = {\rm tr}[\varrho' \Pi_k]$ holds true for all $k$ if and only if $\varrho = \varrho'$, i.e. the quantum state is uniquely determined by the probability distribution of outcomes. Dual operators $\{D_{k'}\}_{k'}$ are defined through the linear inversion formula
\begin{equation}
    \varrho = \sum_k {\rm tr}[\varrho \Pi_k] D_k, \, \forall \varrho
\end{equation}
that relates the density operator $\varrho$ and the probability distribution $\{{\rm tr}[\varrho \Pi_k]\}_k$. 

For example, consider a measuring apparatus that performs a projective measurement in the eigenbasis of one of the Pauli operators $\sigma_x$, $\sigma_y$, $\sigma_z$. If the bases are chosen randomly in accordance with the probability distribution $(p_x,p_y,p_z)$, then 
\begin{eqnarray}
&&   \Pi_{z+} = p_z \ket{0}\bra{0}, \quad \Pi_{z-} = p_z \ket{1}\bra{1}, \\
&&   \Pi_{x\pm} = p_x \ket{\pm}\bra{\pm}, \quad \Pi_{y\pm} = p_y \ket{\pm i}\bra{\pm i},
\end{eqnarray}
where $\{\ket{0},\ket{1}\}$ is the standard computational basis for a qubit ($\sigma_z \ket{0} = \ket{0}$, $\sigma_z \ket{1} = -\ket{1}$), $\ket{\pm} = \frac{1}{\sqrt{2}} (\ket{0} \pm \ket{1})$, and $\ket{\pm i} = \frac{1}{\sqrt{2}} (\ket{0} \pm i \ket{1})$. Since ${\rm dim} \, {\rm Span}(\{\Pi_{k}\}_k) = 4$, the POVM is informationally complete. Physically this corresponds to a possibility of inferring all the Bloch vector components ${\rm tr}[\varrho \sigma_{\alpha}]$, $\alpha = x,y,z$ from the measurement data. The dual operators are not unique in general [for instance, in this case because the number of POVM effects (six) is greater than the dimension of the operator space (four)]. A suitable set of duals is
\begin{equation} \label{duals-example}
    D_{\alpha \pm} = \frac{1}{2}(I \pm p_{\alpha}^{-1} \sigma_{\alpha}).
\end{equation}

Given a quantum register of $N$ qubits, where each qubit is measured individually in the informationally complete way, the linear inversion formula for the whole density operator of all qubits reads
\begin{equation}
    \varrho = \sum_{k_0, \ldots,k_{N-1}} {\rm tr} \left[ \varrho \bigotimes_{m=0}^{N-1} \Pi^{[m]}_{k_m} \right] \ \bigotimes_{m=0}^{N-1} D^{[m]}_{k_m},
\end{equation}
where $\Pi^{[m]}_{k_m}$ and $D^{[m]}_{k_m}$ are the POVM effect and its dual operator for the $m$th qubit, respectively.

\section{Estimation of physical observables with a finite set of measurement outcomes} \label{appendix-estimation}

\begin{figure}
    \centering
    \includegraphics[width = 9cm]{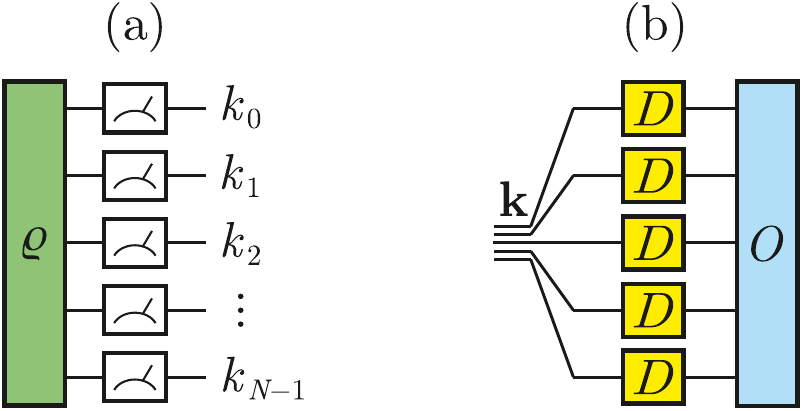}
    \caption{(a) Measurement outcomes for individual qubits form an $N$-tuple $(k_0,\ldots,k_{N-1}) \equiv {\bf k}$. (b) Graphical representation for a real-valued random variable $\xi_{\bf k} := {\rm tr} [H D_{\bf k} ]$ contributing to the estimation $\bar{O} = \frac{1}{S} \sum_{{\bf k} \in {\boldsymbol{\sf S}}} \xi_{\bf k}$ of the physical observable $H$.}
    \label{figure-shots}
\end{figure}

Suppose we run the circuit $S$ times and measure all $N$ qubits individually each time via a fixed informationally complete POVM. Then we get a collection ${\boldsymbol{\sf S}}$ of $S$ measurement outcomes, with each outcome being an $N$-tuple $(k_0,\ldots,k_{N-1}) \equiv {\bf k}$, see Fig.~\ref{figure-shots}(a). The density operator at the circuit output is estimated as $\varrho_{\boldsymbol{\sf S}} = \frac{1}{S} \sum_{{\bf k} \in {\boldsymbol{\sf S}}} D_{\bf k}$, where $D_{\bf k} \equiv \bigotimes_{m=0}^{N-1} D^{[m]}_{k_m}$. For a finite number $S$ of samples, the operator $\varrho_{\boldsymbol{\sf S}}$ may have negative eigenvalues; therefore, we refer to $\varrho_S$ as a quasistate. In the limit of infinitely many measurement outcomes, $\varrho_{\infty} := \lim_{S \rightarrow \infty} \varrho_{\boldsymbol{\sf S}}$ is the true density operator at the output of the quantum circuit. Moreover, the average quasistate is unbiased, $\mathbb{E}[\varrho_{\boldsymbol{\sf S}}] = \varrho$.

Suppose we estimate a physical observable
with the corresponding quantum operator 
$O$, then each measurement outcome ${\bf k}$ induces a real-valued random variable $\xi_{\bf k} := {\rm tr} [D_{\bf k} O]$. The mean $\bar{O} = \frac{1}{S} \sum_{{\bf k} \in {\boldsymbol{\sf S}}} \xi_{\bf k}$ defines an unbiased estimate for the observable, and this estimate is a random variable itself. (The particular realisation for the value of $\bar{O}$ is obtained by running a quantum computation $S$ times.) Since all the random variables $\{\xi_{\bf k}\}_{{\bf k} \in {\boldsymbol{\sf S}}}$ are independent and identically distributed, the variance ${\rm Var}(\bar{O}) = \frac{1}{S} {\rm Var}(\xi)$, where $\xi$ is any of $\xi_{\bf k}$. On the other hand, each ${\rm Var}(\xi)$ can be estimated as $\frac{1}{S} \sum_{{\bf k} \in {\boldsymbol{\sf S}}} (\xi_{\bf k} - \bar{O})^2$, so the final 
statistical error in estimating the observable
reads
\begin{equation} \label{error-formula}
\Delta \bar{O}  = \frac{1}{S} \sqrt{ \sum_{{\bf k} \in {\boldsymbol{\sf S}}} \left( {\rm tr} [D_{\bf k} O] - \bar{O} \right)^2 }.
\end{equation} 

Formula \eqref{error-formula} accounts errors originating from a finite number $S$ of samples (measurements) available in practice. It is equally applicable for both noisy and noiseless circuits, with a difference between the cases being in the set ${\boldsymbol{\sf S}}$ of observed measurement outcomes (${\boldsymbol{\sf S}} = {\boldsymbol{\sf S}}_{\rm noisy}$ or ${\boldsymbol{\sf S}} = {\boldsymbol{\sf S}}_{\rm ideal}$).

In some algorithms, there is an additional level of randomness to be averaged over. For example, in PEC there is averaging over different circuit realisations. In theory, one could collect only one measurement shot per circuit and then formula \eqref{error-formula} would still be valid. This, however, would be impractical in some situations as it could be much faster to collect a number of measurement shots for a single circuit realisation. Suppose we have $M$ measurement shots per circuit and $Q$ circuits, so that the overall budget of shots is $S=QM$. Let $s=(q,m)$ be the multiindex incorporating the circuit number $q$ and the measurement shot number $m$. The corresponding random variable is $\xi(s) := \xi_{{\bf k}(s)} \equiv {\rm tr}[D_{{\bf k}(s)} O]$. Its average value 
\begin{equation} \label{double-average}
    \bar{O} = \frac{1}{QM} \sum_{q,m} \xi(q,m) = \frac{1}{S} \sum_{s} \xi(s)
\end{equation}
gives the unbiased estimation for $\braket{O}$. However, formula \eqref{error-formula} for the error estimation is to be modified, namely, 
\begin{equation} \label{error-formula-settings}
\Delta \bar{O}  = \sqrt{ \frac{1}{(QM)^2} \sum_{q,m} \left[ \xi(q,m) - \xi(q) \right]^2 + \frac{1}{Q^2} \sum_{q} \left[ \xi(q) - \bar{O} \right]^2},
\end{equation} 
where $\xi(q) := \frac{1}{M} \sum_m \xi(q,m)$.

\section{Pauli transfer matrix representation} \label{appendix-PTM}

Consider a linear space of operators acting on the 2-dimensional Hilbert space for a single qubit. Since the identity operator $I \equiv \sigma_0$ and the conventional set of Pauli operators $\sigma_x \equiv \sigma_1$, $\sigma_y \equiv \sigma_2$, $\sigma_z \equiv \sigma_3$ altogether form a basis in this space, any operator $A$ is uniquely determined by a 4-dimensional vector (rank-1 tensor) $\mathfrak{a}$ with components $a_{\alpha} := \frac{1}{\sqrt{2}} {\rm tr}[A \sigma_{\alpha}]$, $\alpha = 0,1,2,3$. The inverse formula reads $A = \frac{1}{\sqrt{2}} \sum_{\alpha = 0}^3 a_\alpha \sigma_\alpha$. The Hilbert-Schmidt scalar product of operators $A$ and $B$ is ${\rm tr}[A^{\dag} B] = \mathfrak{a}^{\dag}\mathfrak{b}$, i.e., corresponds to the conventional scalar product of vectors $\mathfrak{a}$ and $\mathfrak{b}$. 

A linear map ${\cal E}$ on the space of qubit operators is uniquely determined by the $4 \times 4$ matrix (rank-2 tensor) $\mathfrak{E}$ with elements
\begin{equation}
    \mathfrak{E}_{\alpha \beta} := \frac{1}{2} {\rm tr}\left[ \sigma_{\alpha} {\cal E}(\sigma_{\beta}) \right], \quad \alpha,\beta = 0,1,2,3,
\end{equation}
which defines the Pauli transfer matrix (PTM) representation. In the PTM representation, the operator ${\cal E}(A)$ corresponds to the product $\mathfrak{E} \mathfrak{a}$.

A multiqubit generalisation of the PTM representation is straightforward. In the case of $N$ qubits, the operator $A$ corresponds to the vector $a_{\boldsymbol{\alpha}} = \frac{1}{\sqrt{2^N}} {\rm tr}[A \sigma_{\boldsymbol{\alpha}}]$, where $\boldsymbol{\alpha} := (\alpha_0, \ldots, \alpha_{N-1})$ and $\sigma_{\boldsymbol{\alpha}} := \bigotimes_{m=0}^{N-1} \sigma_{\alpha_m}$. The PTM representation for an $N$-qubit map ${\cal E}$ reads $\mathfrak{E}_{\boldsymbol{\alpha} \boldsymbol{\beta}} := \frac{1}{2^N} {\rm tr}\left[ \sigma_{\boldsymbol{\alpha}} {\cal E}(\sigma_{\boldsymbol{\beta}}) \right]$. 

The PTM representation is advantageous as it comes with a straightforward method for constructing composite maps. The PTM form of a composition of two maps (${\cal E} = {\cal F} \circ {\cal G}$) is simply the matrix product of the individual PTM parts ($\mathfrak{E} = \mathfrak{F} \mathfrak{G}$). Similarly, the PTM representation for a tensor product of maps (${\cal E} = {\cal F} \otimes {\cal G}$) is merely a tensor product of the corresponding PTM representations for the maps involved ($\mathfrak{E} = \mathfrak{F} \otimes \mathfrak{G}$). Both properties make the PTM representation ideal for representing a collection of maps acting in succession and potentially on different qubits (such as quantum gates).

\section{Tensor network calculations} \label{appendix-tensor-networks}

Tensor networks provide a computationally efficient description of many quantum-mechanical objects, e.g., the quantum state or a quantum operator. Here we outline how tensor networks can be used to calculate the estimate $\bar{O}$ and its error $\Delta \bar{O}$ for a desired Hermitian operator $H$ based on a given collection ${\boldsymbol{\sf S}}$ of $S$ measurement outcomes.

\subsection{Tensor network for the quasistate} \label{appendix-rho}

\begin{figure}
    \centering
    \includegraphics[width = 12cm]{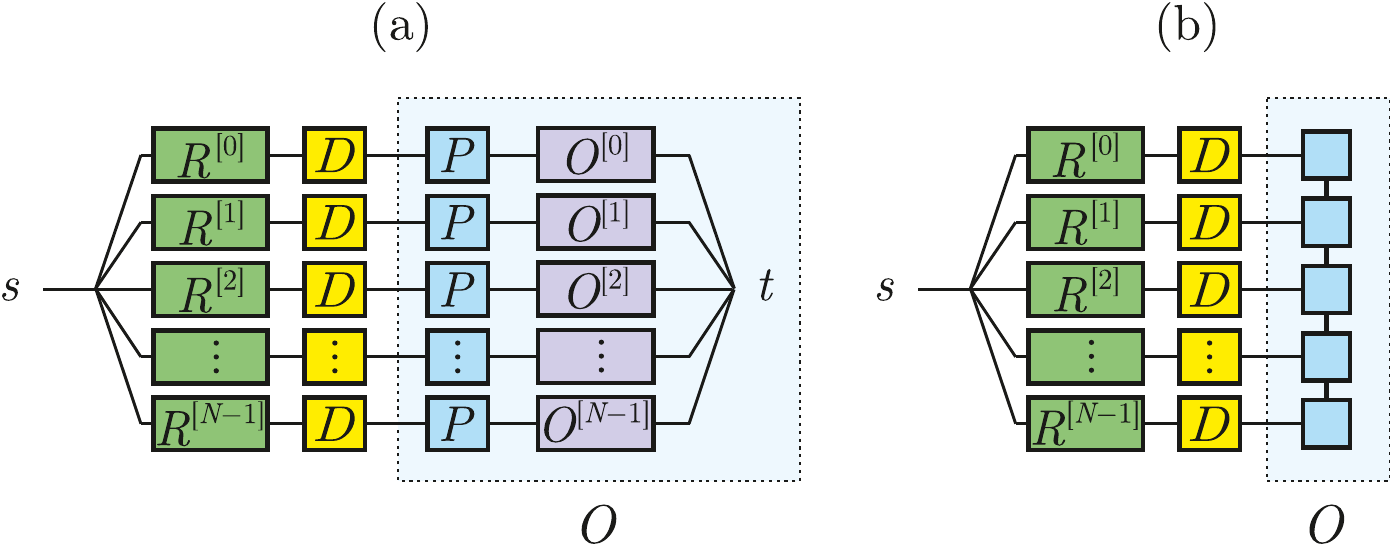}
    \caption{Tensor network for $\xi_{\bf k} \equiv {\rm tr}[D_{\bf k} O]$. The observable $O$ is either the sum of Pauli strings (a) or a matrix product operator (b) in the Pauli transfer matrix representation.}
    \label{figure-observable-t}
\end{figure}

Let us enumerate multiindices ${\bf k}$ in the set ${\boldsymbol{\sf S}}$ by a counter $s=0, \ldots, S-1$, where $S$ is the total number of measurement shots, i.e., ${\boldsymbol{\sf S}} = \{{\bf k}(s)\}_{s = 0}^{S-1}$. Each ${\bf k}(s)$ is a tuple ${\bf (} k_0, \ldots, k_m , \ldots, k_{N-1} {\bf )}$. The set ${\boldsymbol{\sf S}}$ can be viewed as a two-dimensional array of shape $(N,S)$. The quasistate $\varrho_{\boldsymbol{\sf S}} = \frac{1}{S} \sum_{{\bf k} \in {\boldsymbol{\sf S}}} D_{\bf k} = \frac{1}{S} \sum_{{\bf l}} {\bf 1}_{{\boldsymbol{\sf S}}}({\bf l}) D_{\bf l}$, where ${\bf 1}_{{\boldsymbol{\sf S}}}(\cdot)$ is the indicator function for the set ${\boldsymbol{\sf S}}$ and ${\bf l}$ labels all multiqubit POVM elements and their duals (exponentially many in $N$). The indicator function adopts a compact tensor network representation if we consider a collection $\{k_m(s)\}_{s=0}^{S-1}$ of measurement outcomes for each individual qubit number $m$ and introduce the selector matrix $R^{[m]}$ \cite[section 7.2]{boyd-2018} with elements $R^{[m]}_{s l} \in \{0,1\}$ such that $R^{[m]}_{s l} = 1$ if and only if the $s$th multiindex ${\bf k}$ has $m$th component equal to $l$. In terms of the Kronecker delta symbol, $R^{[m]}_{s l} = \delta_{k_m(s),l}$. Then the quasistate takes the form
\begin{equation} \label{quasistate}
    \varrho_{\boldsymbol{\sf S}} = \frac{1}{S} \sum_{s=0}^{S-1} \ \bigotimes_{m = 0}^{N-1} \ \sum_l R^{[m]}_{s l} D_{l}.
\end{equation}
The set of single-qubit dual operators $\{D_l\}_l$ can be considered as a tensor $D$ shown in Fig.~\ref{figure-observable-t}. In the PTM representation, the tensor $D$ has order 2, i.e., it is represented by a matrix. In the case of duals \eqref{duals-example}, the explicit form of this matrix reads
\begin{equation}
    D = \frac{1}{\sqrt{2}} \begin{pmatrix}
    1 & p_x^{-1} & 0 & 0 \\
    1 & -p_x^{-1} & 0 & 0 \\
    1 & 0 & p_y^{-1} & 0 \\
    1 & 0 & -p_y^{-1} & 0 \\
    1 & 0 & 0 & p_z^{-1} \\
    1 & 0 & 0 & -p_z^{-1} \\
    \end{pmatrix},
\end{equation}
where each odd (even) row is a PTM representation of the corresponding dual operator $D_{\alpha +}$ ($D_{\alpha -}$), $\alpha = 1,2,3$. A single term $\bigotimes_{m = 0}^{N-1} \ \sum_l R^{[m]}_{s l} D_{l}$ corresponds to a tensor network on the left from $O$ in Fig.~\ref{figure-observable-t} with a fixed hyperindex $s$. Summing over the hyperindex $s$ and dividing the result by the number of shots, $S$, we get the quasistate \eqref{quasistate}.

\subsection{Tensor network for the observable operator} \label{appendix-H}

The physical observable $O$ is given in the form of an operator acting on the $2^N$-dimensional Hilbert space of $N$ qubits. In quantum chemistry problems, upon utilising a fermion-to-qubit mapping, the operator $O$ is represented as a sum $O = \sum_{\boldsymbol{\alpha}} c_{\boldsymbol{\alpha}} \sigma_{\boldsymbol{\alpha}}$ of Pauli operator strings $\sigma_{\boldsymbol{\alpha}} = \bigotimes_{m=0}^{N-1} \sigma_{\alpha_m}$, where the Pauli operator $\sigma_{\alpha_m}$ acts on the $m$th qubit, $\alpha_m \in \{0,1,2,3\}$. Let us use symbol $t$ to enumerate the multiindices ${\boldsymbol{\alpha}} \equiv (\alpha_0,\ldots,\alpha_{N-1})$ contributing to the sum (i.e., those for which $c_{\boldsymbol{\alpha}} \neq 0$). In typical physical and chemical problems, the number of contributing Pauli strings (dimension of index $t$) is polynomial in the number of qubits $N$ in contrast to the exponentially many contributions for a general observable $O$. Then in full analogy with the quasistate, the operator $O$ adopts the form
\begin{equation} \label{ham-sum}
    O = \sum_{t} \ \bigotimes_{m = 0}^{N-1} \ \sum_{\beta=0}^{3} O^{[m]}_{t \beta} \, \sigma_{\beta},
\end{equation}
where for all $m = 1, \ldots, N-1$ the selector matrix $O^{[m]}$ \cite[section 7.2]{boyd-2018} defines the indicator coefficient $O^{[m]}_{h \beta} = \delta_{\alpha_m(t),\beta}$ such that $O^{[m]}_{t \beta} = 1$ if and only if the $t$th multiindex $\boldsymbol{\alpha}$ has $m$th component equal to $\beta$; whereas for $m = 0$ it also contains the coefficient $c_{\boldsymbol{\alpha}(t)}$ for the $t$th Pauli string, i.e.,  $O^{[0]}_{t \beta} = c_{\boldsymbol{\alpha}(t)} \delta_{\alpha_0(t),\beta}$. The set of single-qubit Pauli operators $\{\sigma_\beta\}_{\beta=0}^3$ can be considered as a `Pauli' tensor $P$ shown in Fig.~\ref{figure-tn-estimations}(a). In the PTM representation, $P$ is a trivial $4 \times 4$ identity matrix multiplied by $\sqrt{2}$, so this tensor can be omitted from the tensor network diagram for the operator $O$ in Fig.~\ref{figure-tn-estimations}(a) with a proper rescaling. Note that the tensor network contains a hyperindex $t$, which can be readily summed over in the popular packages Quimb~\cite{quimb} and Cotengra~\cite{cotengra}. In this sense, the hyperindex $t$ is internal as it is summed over (in contrast to the hyperindex $s$ in the quasistate, which enables us to calculate the error $\Delta \bar{O}$, see explanation in Appendix~\ref{appendix-tn-contraction}). 

Alternatively, the operator $O$ can be originally given in the form of a tensor network, e.g., a well known linear tensor network called the matrix product operator (MPO) \cite{perez-2007,verstraete-2008,schollwock-2011,hubig-2017,montangero-2018,cirac-2021,evenbly-2022}. In the PTM representation, the MPO $O$ takes the form of the unnormalised matrix product state depicted in Fig.~\ref{figure-tn-estimations}(b). This approach to represent the observable is generally more efficient as compared to Eq.~\eqref{ham-sum} because the bond dimension can be generally much less than the number of Pauli strings in $O$.

\subsection{Tensor network contraction} \label{appendix-tn-contraction}

The $s$th measurement shot gives a particular value $\xi(s) := \xi_{{\bf k}(s)} \equiv {\rm tr}[D_{{\bf k}(s)} O]$ for the random variable $\xi_{\bf k}$. This value $\xi(s)$ is exactly the tensor-network contraction shown in Fig.~\ref{figure-tn-estimations} [subfigures (a) and (b) differ in the representation for the operator $O$ only, see Appendix \ref{appendix-H}]. Connected legs indicate indices that are summed over. Contracting either of the tensor networks in Fig.~\ref{figure-tn-estimations} with a fixed value of the outer hyperindex $s$, we get exactly $\xi(s)$. The contraction is routinely performed with the help of packages Quimb~\cite{quimb} and Cotengra~\cite{cotengra} (the latter one finds the optimal contraction tree). 

The estimate $\bar{O}$ for the observable $O$ after $S$ measurement shots is
\begin{equation}
    \bar{O} = \frac{1}{S} \sum_{s = 0}^{S-1} \xi(s).
\end{equation}
The estimation error \eqref{error-formula} reduces to
\begin{equation}
    \Delta\bar{O} = \frac{1}{S} \sqrt{ \sum_{s = 0}^{S-1} \left( \xi(s) - \bar{O} \right)^2}.
\end{equation}

Generalisation to a scenario with double averaging via formulas \eqref{double-average} and \eqref{error-formula-settings} is straightforward.

\begin{figure}
    \centering
    \includegraphics[width = 12cm]{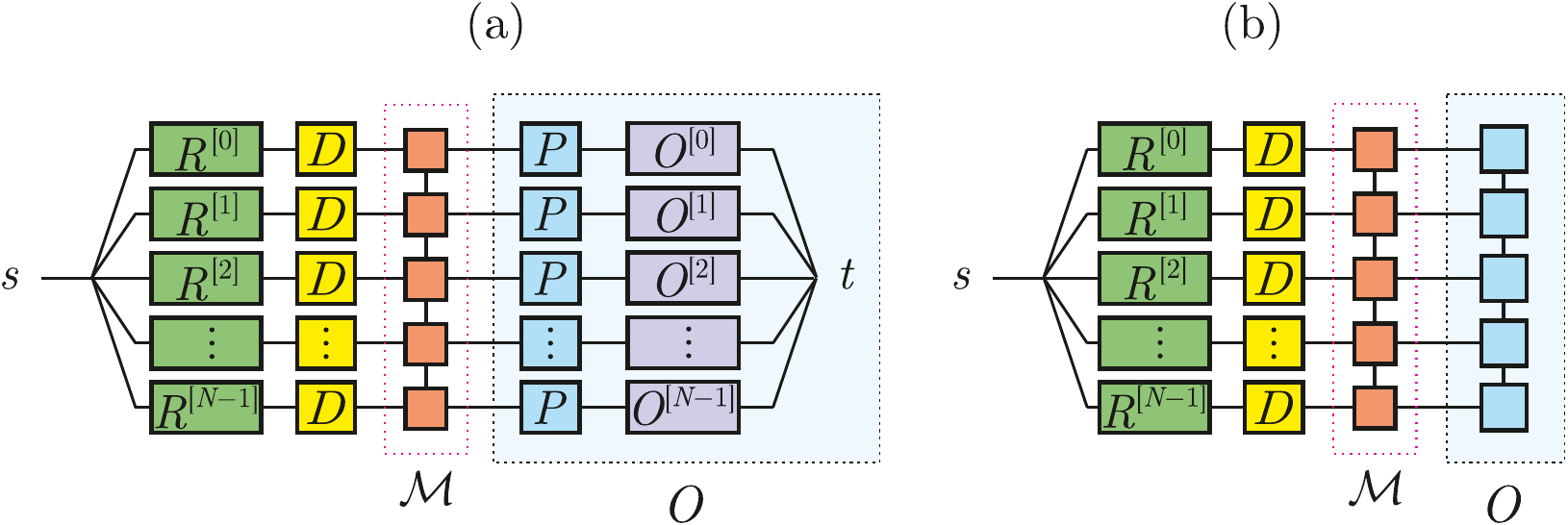}
    \caption{Tensor network for the noise-mitigated estimation $\xi_{\bf k} \equiv {\rm tr}[{\cal M}(D_{\bf k}) O] = \equiv {\rm tr}[D_{\bf k}  {\cal M}^{\dag}(O)]$. The observable $O$ is either the sum of Pauli strings (a) or a matrix product operator (b) in the Pauli transfer matrix representation. ${\cal M}$ is the noise mitigation map having the form of the matrix product operator in the PTM representation.}
    \label{figure-tn-estimations}
\end{figure}

\section{MPO for unitary maps} \label{appendix-mpo-unitary}

Any unitary quantum circuit can be decomposed into single-qubit and two-qubit unitary gates \cite{nielsen-2010}. Moreover, one can restrict the set of two-qubit unitary gates to a single {\sc cnot} gate given by the unitary operator $U_{\text{\sc cnot}} = \ket{0}\bra{0} \otimes I + \ket{1}\bra{1} \otimes \sigma_x$. In the circuit implementation of quantum computation, we can therefore regard a single circuit layer consisting of either single-qubit gates or the {\sc cnot} gate $U_{\text{\sc cnot}}^{[m_1,m_2]}$ acting on qubits $m_1$ and $m_2$ in the register of $N$ qubits.

Consider a layer of single-qubit unitary gates $\{U^{[m]}\}_{m=0}^{N-1}$, where the superscript $m$ indicates the qubit number. Then the unitary map ${\cal U}$ acting on the density operator of the whole register is
\begin{equation}
    {\cal U} = \bigotimes_{m=0}^{N-1} {\cal U}^{[m]},
\end{equation}
where ${\cal U}^{[m]}(\bullet) = U^{[m]} \bullet (U^{[m]})^{\dag}$. In the PTM representation, ${\cal U}$ is an MPO with the trivial bond dimension $1$ (the connecting link is a so called dummy index that takes only one value). Physical input and output for each map ${\cal U}^{[m]}$ have dimension $4$ in the PTM representation. 

Consider a layer consisting of the {\sc cnot} gate $U_{\text{\sc cnot}}^{[m_1,m_2]}$, where $m_1$ is the controlling qubit and $m_2$ is the controlled one. In general, qubits $m_1$ and $m_2$ can be non-adjacent. Suppose $m_1 < m_2$, then the corresponding unitary map for the whole register reads
\begin{equation} \label{U-CNOT}
\sum_{k=0}^3 {\rm Id}^{[0:m_1-1]} \otimes {\cal F}_k^{[m_1]} \otimes {\rm Id}^{[m_1+1:m_2-1]} \otimes {\cal G}_k^{[m_2]} \otimes {\rm Id}^{[m_2+1:N-1]},
\end{equation}
where ${\rm Id}^{[q_1:q_2]} = \bigotimes_{m=q_1}^{q_2} {\rm Id}^{[m]}$ is the identity transformation for qubits with numbers in the range from $q_1$ to $q_2$, $\{{\cal F}_k\}_{k=0}^3$ and $\{{\cal G}_k\}_{k=0}^3$ are collections of qubit maps whose PTM representation reads
\begin{eqnarray}
    && \mathfrak{F}_0 = \left(
\begin{array}{cccc}
 1 & 0 & 0 & 0 \\
 0 & 0 & 0 & 0 \\
 0 & 0 & 0 & 0 \\
 0 & 0 & 0 & 1 \\
\end{array}
\right),\quad \mathfrak{G}_0 = \left(
\begin{array}{cccc}
 1 & 0 & 0 & 0 \\
 0 & 1 & 0 & 0 \\
 0 & 0 & 0 & 0 \\
 0 & 0 & 0 & 0 \\
\end{array}
\right), \quad\\
    && \mathfrak{F}_1 = \left(
\begin{array}{cccc}
 0 & 0 & 0 & 1 \\
 0 & 0 & 0 & 0 \\
 0 & 0 & 0 & 0 \\
 1 & 0 & 0 & 0 \\
\end{array}
\right),\quad \mathfrak{G}_1 = \left(
\begin{array}{cccc}
 0 & 0 & 0 & 0 \\
 0 & 0 & 0 & 0 \\
 0 & 0 & 1 & 0 \\
 0 & 0 & 0 & 1 \\
\end{array}
\right),\quad\\
    && \mathfrak{F}_2 = \left(
\begin{array}{cccc}
 0 & 0 & 0 & 0 \\
 0 & 1 & 0 & 0 \\
 0 & 0 & 1 & 0 \\
 0 & 0 & 0 & 0 \\
\end{array}
\right),\quad \mathfrak{G}_2 = \left(
\begin{array}{cccc}
 0 & 1 & 0 & 0 \\
 1 & 0 & 0 & 0 \\
 0 & 0 & 0 & 0 \\
 0 & 0 & 0 & 0 \\
\end{array}
\right),\quad\\
    && \mathfrak{F}_3 = \left(
\begin{array}{cccc}
 0 & 0 & 0 & 0 \\
 0 & 0 & 1 & 0 \\
 0 & -1 & 0 & 0 \\
 0 & 0 & 0 & 0 \\
\end{array}
\right),\quad \mathfrak{G}_3 = \left(
\begin{array}{cccc}
 0 & 0 & 0 & 0 \\
 0 & 0 & 0 & 0 \\
 0 & 0 & 0 & 1 \\
 0 & 0 & -1 & 0 \\
\end{array}
\right).\qquad
\end{eqnarray}
In the PTM representation, the unitary map \eqref{U-CNOT} is given by the MPO with a varying bond dimension: the bond dimension equals $1$ (dummy index) for links between qubits $0$ and $m_1$, $m_2$ and $N-1$; other links have bond dimension $4$. The final MPO for \eqref{U-CNOT} in the PTM representation is
\begin{equation}
    \sum_{a_0,\ldots,a_{N-2}} \mathfrak{E}^{[0]}_{a_0} \otimes \mathfrak{E}^{[1]}_{a_0 a_1} \otimes \cdots \otimes \mathfrak{E}^{[N-2]}_{a_{N-3} a_{N-2}} \otimes \mathfrak{E}^{[N-1]}_{a_{N-2}},
\end{equation}
where $a_0 = \ldots = a_{m_1-1} = a_{m_2} = \ldots = a_{N-2} = 0$ (dummy indices) and $\mathfrak{E}_{0}^{[0]} = \ldots = \mathfrak{E}_{00}^{[m_1-1]} = \mathfrak{E}_{00}^{[m_2+1]} = \ldots = \mathfrak{E}_{0}^{[N-1]} = I_{4 \times 4}$ for the corresponding qubits; $a_{m_1} = \ldots = a_{m_2 - 1} \in \{0,1,2,3\}$ and for the {\sc cnot}-involved qubits $m_1$ and $m_2$ we have $\mathfrak{E}_{0 a_{m_1}}^{[m_1]} = {F}_{a_{m_1}}$ and $\mathfrak{E}_{a_{m_2-1}0}^{[m_2]} = \mathfrak{G}_{a_{m_2-1}}$, whereas for the intermediate qubits $q$ in the range from $m_1+1$ to $m_2-1$ we have $\mathfrak{E}_{a_{q-1} a_{q}}^{[q]} = \delta_{a_{q-1} a_{q}} I_{4 \times 4}$. 

If the circuit is composed of $k$-local unitary gates other than {\sc cnot}, then the PTM representation of the corresponding unitary maps can be routinely transformed into an MPO by using a general decomposition procedure [e.g., the singular value decomposition (SVD) described in Ref.~\cite{schollwock-2011}]. In general, this method takes some rectangular matrix $A$ of shape $(n \times p)$, and decomposes it into $ A_{n \times p} = U_{n \times n} \Sigma_{n \times p} V _{p \times p} ^{\dagger} $, where the columns of $U$ are the left singular vectors, $\Sigma$ has the singular values of $A$ along its diagonal, and $V ^{\dagger}$ has rows that are the right singular vectors. Let us illustrate this with a $2$-local unitary map with the PTM representation $\mathfrak{A}^{(i_1,i_2)}_{(o_1, o_2)}$, where $(i_1,i_2)$ are input indices and  ${(o_1, o_2)}$ are output indices corresponding to qubits $1$ and $2$.  If we were to reorder the indices of $\mathfrak{A}^{(i_1,i_2)}_{(o_1, o_2)}$ to give $\mathfrak{A}^{(i_1,o_1)}_{(i_2, o_2)}$, and then perform the SVD with respect to multiindices ${(i_1,o_1)}$ on one side and ${(i_2, o_2)}$  on the other side, we would have 
$\mathfrak{A}^{(i_1,o_1)}_{(i_2,o_2)} = U^{(i_1,o_1)}_{ \mu } \Sigma ^{\mu} _{ \mu '} (V^{\dagger}) ^{\mu '} _{(i_2,o_2)} $ with individual tensors $U ^{(i_1,o_1)}_{\mu}$ and $\Sigma ^{\mu}_{\mu'} (V^{\dagger})^{\mu'}_{(i_2,o_2)}$ on each qubit connected by some index $\mu$. The bond dimension $\{\mu\}$ does not exceed $16$ in this case. After performing this decomposition on each 2-local map, we will have a tensor network with connecting links between single-qubit maps, i.e., the MPO form for each unitary layer in the circuit. A generalisation of this method to $k$-local unitary gates involves $k-1$ decompositions and follows the lines of constructing the MPO for a given operator.

Generally, to allow for the fact that the whole circuit may be relatively deep, it can be segmented into individual subcircuits of shallow depth, where each subcircuit admits a small number of gates. After each $k$-local unitary map in the circuit is decomposed, the MPO form appears naturally by contracting any ``horizontal'' index with respect to the direction of the circuit. In our implementation, the subcircuits consist of individual layers (single qubit gates or non-overlapping {\sc cnot} gates), and each such layer is transformed into a simple MPO in the PTM representation (with bond dimension $1$ or $4$) as described above in this section. 

\section{Noise inversion map and its MPO form} \label{appendix-noise-mpo}

As a consequence of noise, the true physical implementation of each unitary map $\mathcal{U}$ is some noisy quantum channel $\mathcal{E} \equiv \mathcal{N} \circ \mathcal{U} \neq \mathcal{U}$. Suppose this noisy channel is fully characterised with a reasonable accuracy, e.g., via the process tomography \cite{nielsen-2021}. Then the noisy map $\mathcal{N} = \mathcal{E} \circ \mathcal{U}^{-1}$. If the gate is implemented with a high fidelity, then $\mathcal{E} \approx \mathcal{U}$ and $\mathcal{N} \approx {\rm Id}$. Due to the latter fact, $\mathcal{N}^{-1}$ is always well defined whenever the noise level reasonably small. In the PTM representation, the inverse noise map is defined by the matrix $\mathfrak{N}^{-1} = \mathfrak{U}\mathfrak{E}^{-1}$. Assuming the gates act locally on a few qubits, all the matrix operations are readily implementable. The map $\mathcal{N}^{-1}$ is then represented in the MPO form in full analogy with unitary maps (see Appendix~\ref{appendix-mpo-unitary}). In Sections \ref{section-two-qubit-depolarizing} and \ref{section-global-depolarizing}, we show that the MPO has bond dimension 2 in the case of depolarising noise. In Sec.~\ref{section-two-qubit-pauli-channel}, we consider a general Pauli qubit noise affecting 2 qubits and show that the corresponding MPO has bond dimension 4. 

In actual hardware, noise affects not only qubits subjected to a local unitary gate. Nearby qubits are vulnerable to the unavoidable crosstalk. Additionally, idle qubits decohere too. Therefore, a more general noise model should take those effects into account. On the other hand, the model should be scalable and avoid the exponentially heavy tomography. A recently studied sparse Pauli-Lindblad model provides an effective noise description in actual devices exploiting the randomised compiling \cite{berg-2023}. In the case of the linear topology for an $N$-qubit register, the model contains $12N-9$ parameters [$3N$ of which describe single-qubit decoherence rates and $9(N-1)$ are associated with the nearest-qubit crosstalk decoherence rates]. The parameters can be learned with the near-constant learning cost in $N$ \cite{berg-2023}. In Sec.~\ref{appendix-pauli-lindblad-noise-model}, we construct a concise tensor network description for that model in terms of the MPO with the bond dimension 4.

\subsection{2-qubit depolarising noise} \label{section-two-qubit-depolarizing}
Let the noisy map $\mathcal{N}$ be a two-qubit depolarising map with the noise intensity $\epsilon$, i.e., 
\begin{equation} \label{2-qubit-dep}
    \mathcal{N}[\varrho] = (1-\epsilon) \varrho + \epsilon {\rm tr}[\varrho] \frac{1}{4} I_{4 \times 4}.
\end{equation} 
Then the noise-inversion map reads 
\begin{equation*} \label{2-qubit-dep-inverse}
    \mathcal{N}^{-1}[\varrho] = \frac{1}{1-\epsilon} \varrho - \frac{\epsilon}{1-\epsilon} {\rm tr}[\varrho] \frac{1}{4} I_{4 \times 4}. 
\end{equation*} 
The map $\mathcal{N}^{-1}$ has the interqubit bond dimension $2$ if $\epsilon \in (0,1)$. This follows from the fact that 
\begin{equation*}
    \mathcal{N}^{-1} = \mathcal{F}_0 \otimes \mathcal{G}_0 + \mathcal{F}_1 \otimes \mathcal{G}_1,
\end{equation*} 
where $\mathcal{F}_0 = \mathcal{G}_0 = (1-\epsilon)^{-1/2} {\rm Id}$ is a rescaled identity map for a single qubit and $\mathcal{F}_1 [\bullet] = - \mathcal{G}_1 [\bullet] = \sqrt{\epsilon(1-\epsilon)^{-1}} {\rm tr}[\bullet] \frac{1}{2} I_{2 \times 2}$. 

Reshaping the $16 \times 16$ PTM representation $(\mathfrak{N}^{-1})^{(i_1,i_2)}_{(o_1, o_2)}$ for the map $\mathcal{N}^{-1}$ into $(\mathfrak{N}^{-1})^{(i_1,o_1)}_{(i_2, o_2)}$, where $(i_m,o_m)$ is the input-output multiindex for the $m$th qubit, we explicitly find non-zero singular values with respect to the interqubit link: 
\begin{eqnarray}
    \Sigma_0^0 &=& \frac{\sqrt{16-2 \epsilon+\epsilon^2+(4-\epsilon ) \sqrt{16+4 \epsilon + \epsilon^2}}}{\sqrt{2} (1-\epsilon )} \nonumber\\
    &\approx & 4 \left(1+\frac{15\epsilon}{16}\right), \label{sv-0-dep} \\
    \Sigma_1^1 &=& \frac{\sqrt{16-2 \epsilon+\epsilon^2-(4-\epsilon ) \sqrt{16+4 \epsilon + \epsilon^2}}}{\sqrt{2} (1-\epsilon )} \nonumber\\
    &\approx & \frac{3\epsilon}{4}.
\end{eqnarray}

If $\epsilon = 0$, then $\mathcal{N}^{-1} = {\rm Id}$ and we are left with the only non-zero singular value $\Sigma_0^0 = 4$. Deviation of this singular value from $1$ is not surprising as we decompose the map with respect to qubits [${(i_1,o_1)}$ vs ${(i_2, o_2)}$], not with respect to input and output [${(i_1,i_2)}$ vs ${(o_1, o_2)}$]. The physical meaning of the leading singular value \eqref{sv-0-dep} becomes clear if we consider the energy functional ${\rm tr}\left[\mathcal{N}^{-1}[\varrho] H\right] = {\rm tr}\left[\varrho \mathcal{N}^{-1}[H]\right]$. The Pauli string expansion for $H = \sum_{\bf l} c_{\bf l} P_{\bf l}$ after application of the noise-inversion map $\mathcal{N}^{-1}$ takes the form $\mathcal{N}^{-1}[H] = \sum_{\bf l} c_{\bf l}' P_{\bf l}$, where $c_{\bf l}' = c_{\bf l}$ if the noise-affected substring of $P_{\bf l}$ equals $I \otimes I$, $c_{\bf l}' = (1-\epsilon)^{-1} c_{\bf l}$ if the noise-affected substring of $P_{\bf l}$ differs from $I \otimes I$ (15 different possibilities: $I\otimes X$, \ldots, $Z \otimes Z$). If we assume that all Pauli strings have similar contributions to the observable and appear with the same frequency, then on average $c_{\bf l}' \approx (\frac{1}{16} \times 1 + \frac{15}{16} \times (1-\epsilon)^{-1}) c_{\bf l} \approx \left(1+\frac{15\epsilon}{16}\right) c_{\bf l}$. These arguments are directly applicable to the estimation of the measurement overhead (see Appendix~\ref{appendix-measurement-overhead}).

\subsection{Global depolarising noise} \label{section-global-depolarizing}
Suppose the noisy map $\mathcal{N}$ is an $N$-qubit global depolarising channel with the noise intensity $\epsilon$, i.e., 
\begin{equation*}
\mathcal{N}[\varrho] = (1-\epsilon) \varrho + \epsilon {\rm tr}[\varrho] \frac{1}{2^N} I_{2^N \times 2^N}.
\end{equation*}
Then the noise-inversion map reads 
\begin{equation*}
    \mathcal{N}^{-1}[\varrho] = \frac{1}{1-\epsilon} \varrho - \frac{\epsilon}{1-\epsilon} {\rm tr}[\varrho] \frac{1}{2^N} I_{2^N \times 2^N}.
\end{equation*}
The map $\mathcal{N}^{-1}$ has the bond dimension $2$ if $\epsilon \in (0,1)$ because $\mathcal{N}^{-1} = \mathcal{F}_0^{\otimes N} + \mathcal{F}_1^{\otimes N}$, where $\mathcal{F}_0 = (1-\epsilon)^{-1/N} {\rm Id}$ is a rescaled identity map for a single qubit and $\mathcal{F}_1 [\bullet] =  [-\epsilon(1-\epsilon)^{-1}]^{1/N} {\rm tr}[\bullet] \frac{1}{2} I_{2 \times 2}$. In the MPO for $\mathcal{N}^{-1}$, non-zero contributions are only those where the virtual indices are either all equal to 0 or all equal to 1 (like in the matrix product representation for the Greenberger--Horne--Zeilinger state).

If only global depolarizing noise is present in the quantum circuit, then the calculation of the noise mitigation map is trivial because $\mathcal{N}^{-1}$ commutes with any unitary operation ${\cal U}$. In the case of $L$ noisy layers, we have 
\begin{eqnarray*}
    {\cal M}[\varrho] &=& \mathcal{N}^{-L}[\varrho] \\
    &=& \frac{1}{(1-\epsilon)^L} \varrho + \left( 1 - \frac{1}{(1-\epsilon)^L} \right) {\rm tr}[\varrho] \frac{1}{2^N} I_{2^N \times 2^N},
\end{eqnarray*}
so ${\cal M}$ also has the bond dimension 2. Therefore, the bond dimension $\chi_{\rm max} = 2$ suffices to fully mitigate the global depolarising noise without any compression error. This observation is aligned with the simple mitigation of global depolarising errors proposed in Ref.~\cite{vovrosh-2021}.

\subsection{2-qubit Pauli noise} \label{section-two-qubit-pauli-channel}
Let the noisy map $\mathcal{N}$ be a two-qubit Pauli channel. In the PTM representation, $\mathfrak{N} = {\rm diag}(1,\varkappa_{01},\ldots,\varkappa_{33})$, where $15$ real parameters $\varkappa_{ij}$ define the scaling coefficients for the operators $\sigma_i \otimes \sigma_j$. Assuming the noise intensity is relatively small, $\varkappa_{ij} = 1 - \epsilon_{ij}$, where $0 \leq \epsilon_{ij} \ll 1$. The inverse map $\mathcal{N}^{-1}$ generally has the interqubit bond dimension $4$ because the PTM representation $\mathfrak{N}^{-1} = {\rm diag}(1,\varkappa_{01}^{-1},\ldots,\varkappa_{33}^{-1})$ adopts the decomposition $\mathfrak{N}^{-1} = \sum_{i=0}^3 \mathfrak{F}_i \otimes \mathfrak{G}_i$, where $\mathfrak{F}_i = {\rm diag}(\delta_{i0},\delta_{i1},\delta_{i2},\delta_{i3})$ and $\mathfrak{G}_i = {\rm diag}(\varkappa_{i0}^{-1},\varkappa_{i1}^{-1},\varkappa_{i2}^{-1},\varkappa_{i3}^{-1})$. Reshaping the $16 \times 16$ PTM representation $(\mathfrak{N}^{-1})^{(i_1,i_2)}_{(o_1, o_2)}$ for the map $\mathcal{N}^{-1}$ into $(\mathfrak{N}^{-1})^{(i_1,o_1)}_{(i_2, o_2)}$, where $(i_m,o_m)$ is the input-output multiindex for the $m$th qubit, we can explicitly find non-zero singular values with respect to the interqubit link. The largest singular value in the first order of the error parameters reads
\begin{equation}
    \Sigma_0^0 \approx 4 \left( 1 + \frac{1}{16} \sum_{ij} \epsilon_{ij} \right).
\end{equation}

Similarly to the case of depolarising noise, one can interpret the quarter of this singular value as the average multiplicative factor in estimating a typical observable. To recapitulate, we consider the functional ${\rm tr}\left[\mathcal{N}^{-1}[\varrho] H\right] = {\rm tr}\left[\varrho \mathcal{N}^{-1}[H]\right]$ for the observable $H$. The Pauli string expansion for $H = \sum_{\bf l} c_{\bf l} P_{\bf l}$ after application of the noise-inversion map $\mathcal{N}^{-1}$ takes the form $\mathcal{N}^{-1}[H] = \sum_{\bf l} c_{\bf l}' P_{\bf l}$, where $c_{\bf l}' = (1-\epsilon_{ij})^{-1} c_{\bf l}$ if the noise-affected substring of $P_{\bf l}$ equals $\sigma_i \otimes \sigma_j$. If we assume that all Pauli strings have similar contributions to the observable $H$ and appear with the same frequency, then on average $c_{\bf l}' \approx \left( 1 + \frac{1}{16} \sum_{ij} \epsilon_{ij} \right) c_{\bf l}$. These arguments are again directly applicable to the estimation of the measurement overhead (see Appendix~\ref{appendix-measurement-overhead}).

\subsection{Sparse Pauli-Lindblad noise model} \label{appendix-pauli-lindblad-noise-model}

\begin{figure}
    \centering
    \includegraphics[width = 8.5cm]{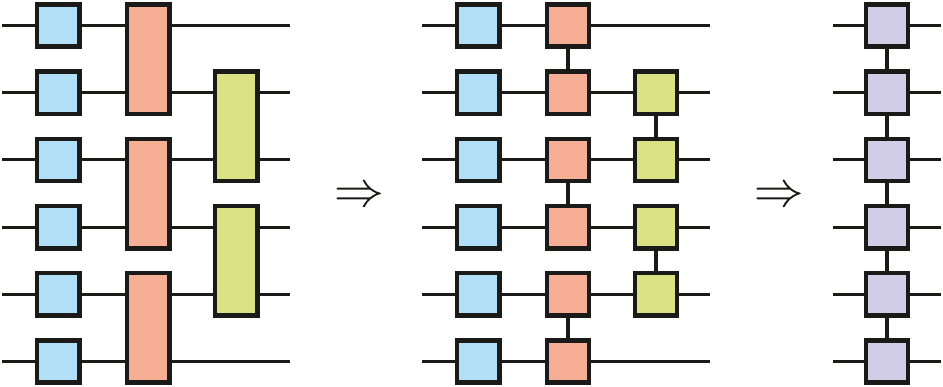}
    \caption{MPO construction for the inverse of the sparse Pauli-Lindblad noise model (linear topology with the nearest neighbour crosstalk). Single-qubit maps and two-qubit maps in Eq.~\eqref{inverse-sparse-pauli-lindblad} commute. Each of the two-qubit maps adopts a decomposition with the bond dimension $4$. Horizontal contractions result in the MPO representation.}
    \label{figure-sparce-pauli-lindblad-tn}
\end{figure}

Consider an $N$-qubit Pauli string $\sigma_{\boldsymbol{\alpha}} = \bigotimes_{m=0}^{N-1} \sigma_{\alpha_m}$ as a jump operator in the Lindblad superoperator $L_{\boldsymbol{\alpha}}(\bullet) = \lambda_{\boldsymbol{\alpha}}(\sigma_{\boldsymbol{\alpha}} \bullet \sigma_{\boldsymbol{\alpha}} - \bullet)$ with the rate $\lambda_{\boldsymbol{\alpha}} \geq 0$. Note that these Lindblad superoperators commute, i.e., $L_{\boldsymbol{\alpha}}\left(L_{\boldsymbol{\beta}}(\bullet)\right) = L_{\boldsymbol{\beta}}\left(L_{\boldsymbol{\alpha}}(\bullet)\right)$. This implies the Pauli channel expansion
\begin{equation}
    {\cal N} \equiv \exp\left( \sum_{\boldsymbol{\alpha}} L_{\boldsymbol{\alpha}} \right) = \bigcirc_{\boldsymbol{\alpha}} e^{ L_{\boldsymbol{\alpha}} }.
\end{equation}
The expansion is particularly useful in the case of the local noise, for which each map $e^{ L_{\boldsymbol{\alpha}} }$ acts trivially on all but potentially a few adjacent qubits. Restricting to the single- and two-qubit local maps, we get the sparse model with $3N+9(N-1)$ potentially non-zero parameters $\lambda_{\boldsymbol{\alpha}}$. We regroup the maps $e^{ L_{\boldsymbol{\alpha}} }$ according to the location of their non-trivial action, namely,
\begin{equation}
    {\cal N} = \left( \prod\limits_{m=0}^{N-1} {\cal N}^{[m]} \right) \circ \left( \prod\limits_{m=0}^{N-2} {\cal N}^{[m,m+1]} \right),
\end{equation}
where
\begin{eqnarray*}
    && {\cal N}^{[m]} = \exp \left( \sum_{\alpha_m = 1}^{3} L_{(0,\ldots,0,\alpha_m,0,\ldots,0)} \right), \nonumber\\
    && {\cal N}^{[m,m+1]} = \exp \left( \sum_{\alpha_m,\alpha_{m+1} = 1}^{3} L_{(0,\ldots,0,\alpha_m,\alpha_{m+1},0,\ldots,0)} \right). 
\end{eqnarray*}
${\cal N}^{[m]}$ acts non-trivially at the $m$th qubit only, so in what follows it will be considered as the single-qubit map with 3 parameters $\{\lambda_i^{[m]}\}_{i=1}^3$. ${\cal N}^{[m,m+1]}$ acts non-trivially at the $m$th and $(m+1)$st qubits only, so in what follows it will be considered as the two-qubit map with 9 parameters $\{\lambda_{ij}^{[m]}\}_{i,j=1}^{3}$. The straightforward calculation yields the diagonal PTM representation for each of the maps, namely, $\mathfrak{N}^{[m]}  = {\rm diag}(1,\varkappa_1^{[m]},\varkappa_2^{[m]},\varkappa_3^{[m]})$ and $\mathfrak{N}^{[m,m+1]}  = {\rm diag}(1,\varkappa_{11}^{[m]},\ldots,\varkappa_{33}^{[m]})$ with
\begin{eqnarray*}
    && \!\!\!\!\! \varkappa_1^{[m]} = \exp\left(-2(\lambda_2^{[m]} + \lambda_3^{[m]})\right),\\
    && \!\!\!\!\! \varkappa_2^{[m]} = \exp\left(-2(\lambda_3^{[m]} + \lambda_1^{[m]})\right),\\
    && \!\!\!\!\! \varkappa_3^{[m]} = \exp\left(-2(\lambda_1^{[m]} + \lambda_2^{[m]})\right),\\
    && \!\!\!\!\! \varkappa_{01}^{[m]} = \exp\left(-2(\lambda_{12}^{[m]} \!+\! \lambda_{13}^{[m]} \!+\! \lambda_{22}^{[m]} \!+\! \lambda_{23}^{[m]} \!+\! \lambda_{32}^{[m]} \!+\! \lambda_{33}^{[m]} )\right),\\
    && \!\!\!\!\! \varkappa_{02}^{[m]} = \exp\left(-2(\lambda_{11}^{[m]} \!+\! \lambda_{13}^{[m]} \!+\! \lambda_{21}^{[m]} \!+\! \lambda_{23}^{[m]} \!+\! \lambda_{31}^{[m]} \!+\! \lambda_{33}^{[m]} )\right),\\
    && \!\!\!\!\! \varkappa_{03}^{[m]} = \exp\left(-2(\lambda_{11}^{[m]} \!+\! \lambda_{12}^{[m]} \!+\! \lambda_{21}^{[m]} \!+\! \lambda_{22}^{[m]} \!+\! \lambda_{31}^{[m]} \!+\! \lambda_{32}^{[m]} )\right),\\
    && \!\!\!\!\! \varkappa_{10}^{[m]} = \exp\left(-2(\lambda_{21}^{[m]} \!+\! \lambda_{22}^{[m]} \!+\! \lambda_{23}^{[m]} \!+\! \lambda_{31}^{[m]} \!+\! \lambda_{32}^{[m]} \!+\! \lambda_{33}^{[m]} )\right),\\
    && \!\!\!\!\! \varkappa_{11}^{[m]} = \exp\left(-2(\lambda_{12}^{[m]} \!+\! \lambda_{13}^{[m]} \!+\! \lambda_{21}^{[m]} \!+\! \lambda_{31}^{[m]} )\right),\\
    && \!\!\!\!\! \varkappa_{12}^{[m]} = \exp\left(-2(\lambda_{11}^{[m]} \!+\! \lambda_{13}^{[m]} \!+\! \lambda_{22}^{[m]} \!+\! \lambda_{32}^{[m]} )\right),\\
    && \!\!\!\!\! \varkappa_{13}^{[m]} = \exp\left(-2(\lambda_{11}^{[m]} \!+\! \lambda_{12}^{[m]} \!+\! \lambda_{23}^{[m]} \!+\! \lambda_{33}^{[m]} )\right),\\
    && \!\!\!\!\! \varkappa_{20}^{[m]} = \exp\left(-2(\lambda_{11}^{[m]} \!+\! \lambda_{12}^{[m]} \!+\! \lambda_{13}^{[m]} \!+\! \lambda_{31}^{[m]} \!+\! \lambda_{32}^{[m]} \!+\! \lambda_{33}^{[m]} )\right),\\
    && \!\!\!\!\! \varkappa_{21}^{[m]} = \exp\left(-2(\lambda_{11}^{[m]} \!+\! \lambda_{22}^{[m]} \!+\! \lambda_{23}^{[m]} \!+\! \lambda_{31}^{[m]} )\right),\\
    && \!\!\!\!\! \varkappa_{22}^{[m]} = \exp\left(-2(\lambda_{12}^{[m]} \!+\! \lambda_{21}^{[m]} \!+\! \lambda_{23}^{[m]} \!+\! \lambda_{32}^{[m]} )\right),\\
    && \!\!\!\!\! \varkappa_{23}^{[m]} = \exp\left(-2(\lambda_{13}^{[m]} \!+\! \lambda_{21}^{[m]} \!+\! \lambda_{22}^{[m]} \!+\! \lambda_{33}^{[m]} )\right),\\
    && \!\!\!\!\! \varkappa_{30}^{[m]} = \exp\left(-2(\lambda_{11}^{[m]} \!+\! \lambda_{12}^{[m]} \!+\! \lambda_{13}^{[m]} \!+\! \lambda_{21}^{[m]} \!+\! \lambda_{22}^{[m]} \!+\! \lambda_{23}^{[m]} )\right),\\
    && \!\!\!\!\! \varkappa_{31}^{[m]} = \exp\left(-2(\lambda_{11}^{[m]} \!+\! \lambda_{21}^{[m]} \!+\! \lambda_{32}^{[m]} \!+\! \lambda_{33}^{[m]} )\right),\\
    && \!\!\!\!\! \varkappa_{32}^{[m]} = \exp\left(-2(\lambda_{12}^{[m]} \!+\! \lambda_{22}^{[m]} \!+\! \lambda_{31}^{[m]} \!+\! \lambda_{33}^{[m]} )\right),\\
    && \!\!\!\!\! \varkappa_{33}^{[m]} = \exp\left(-2(\lambda_{13}^{[m]} \!+\! \lambda_{23}^{[m]} \!+\! \lambda_{31}^{[m]} \!+\! \lambda_{32}^{[m]} )\right).
\end{eqnarray*}

The inverse map
\begin{equation} \label{inverse-sparse-pauli-lindblad}
    {\cal N}^{-1} = \left( \prod\limits_{m=0}^{N-1} ({\cal N}^{[m]})^{-1}  \right) \circ \left( \prod\limits_{m=0}^{N-2} ({\cal N}^{[m,m+1]})^{-1}  \right)
\end{equation}
is obtained from ${\cal N}$ by changing sign of all $\lambda$-parameters. Commutativity of maps $({\cal N}^{[m,m+1]})^{-1}$ makes it possible to consider $\prod\limits_{m=0}^{N-2} ({\cal N}^{[m,m+1]})^{-1}$ as a single brick-wall layer $ \prod\limits_{m \text{~is~even}} ({\cal N}^{[m,m+1]})^{-1} \circ \prod\limits_{m \text{~is~odd}} ({\cal N}^{[m,m+1]})^{-1}$. Each map $({\cal N}^{[m,m+1]})^{-1}$ is a two-qubit Pauli map adopting an MPO form $({\cal N}^{[m,m+1]})^{-1} = \sum_{a_{m+1}=0}^3 {\cal F}_{a_{m+1}}^{[m]} \otimes {\cal G}_{a_{m+1}}^{[m+1]}$ with the bond dimension 4 (see Sec.~\ref{section-two-qubit-pauli-channel}). Merging single-qubit maps into
\begin{equation*}
    {\cal A}_{a_m,a_{m+1}}^{[m]} = ({\cal N}^{[m]})^{-1} \circ {\cal G}_{a_m}^{[m]} \circ {\cal F}_{a_{m+1}}^{[m]},
\end{equation*}
we get the MPO representation 
\begin{equation}
    {\cal N}^{-1} = \sum_{a_0,\ldots,a_{N-2}=0}^{3} {\cal A}^{[0]}_{a_0} \otimes {\cal A}^{[1]}_{a_0 a_1} \otimes \cdots \otimes {\cal A}^{[N-2]}_{a_{N-3} a_{N-2}} \otimes {\cal A}^{[N-1]}_{a_{N-2}} 
\end{equation}
with the bond dimension $\chi_{\rm n} = 4$ (see Fig.~\ref{figure-sparce-pauli-lindblad-tn} for the graphical explanation of the MPO construction).

\section{Noise mitigation map as an MPO}

This section is devoted to details behind the iterative construction of the noise mitigation map via Eq.~\eqref{middle-out-iteration} in the main text. The multiplication and compression of MPOs are well known \cite{perez-2007,verstraete-2008,schollwock-2011,hubig-2017,montangero-2018,cirac-2021,evenbly-2022} and routinely implemented in popular computation packages, e.g., in Quimb \cite{quimb}, here we review them for the sake of completeness. 

\subsection{Multiplication of MPOs} \label{appendix-multiplication-mpos}
Multiplying two matrix product operators for $N$ subsystems
\begin{equation*}
    \mathfrak{A} = \sum_{a_0,\ldots,a_{N-2}} \mathfrak{A}_{a_0}^{[0]} \otimes \mathfrak{A}_{a_0 a_1}^{[1]} \otimes \cdots \otimes \mathfrak{A}_{a_{N-3} a_{N-2}}^{[N-2]} \otimes \mathfrak{A}_{a_{N-2}}^{[N-1]}, 
\end{equation*}
\begin{equation*}
    \mathfrak{B} = \sum_{b_0,\ldots,b_{N-2}} \mathfrak{B}_{b_0}^{[0]} \otimes \mathfrak{B}_{b_0 b_1}^{[1]} \otimes \cdots \otimes \mathfrak{B}_{b_{N-3} b_{N-2}}^{[N-2]} \otimes \mathfrak{B}_{b_{N-2}}^{[N-1]},
\end{equation*}
we get another operator $\mathfrak{C} = \mathfrak{A} \mathfrak{B}$ in the MPO form
\begin{equation}
    \mathfrak{C} = \sum_{c_0,\ldots,c_{N-2}} \mathfrak{C}_{c_0}^{[0]} \otimes \mathfrak{C}_{c_0 c_1}^{[1]} \otimes \cdots \otimes \mathfrak{C}_{c_{N-3} c_{N-2}}^{[N-2]} \otimes \mathfrak{C}_{c_{N-2}}^{[N-1]},
\end{equation}
where the virtual index $c_m = (a_m b_m)$ is the multiindex composed of virtual indices $a_m$ and $b_m$ so that the bond dimension $|\{c_m\}| = |\{a_m\}| \cdot |\{b_m\}|$, and the operator $\mathfrak{C}_{c_{m-1} c_m}^{[m]} = \mathfrak{A}_{a_{m-1} a_m}^{[m]} \mathfrak{B}_{b_{m-1} b_m}^{[m]}$ \cite{perez-2007,verstraete-2008,schollwock-2011,hubig-2017,montangero-2018,cirac-2021,evenbly-2022}.

\subsection{MPO compression} \label{appendix-compression}

Suppose we have an MPO $\mathfrak{A}$ with bond dimension $\chi$ for $N$ subsystems and we want to approximate it by another MPO $\mathfrak{B}$ with a smaller bond dimension $\chi_{\rm max}$. Then the standard procedure would be to bring $\mathfrak{A}$ to a canonical form and leave the most contributing $\chi_{\rm max}$ singular values $\{\lambda_i\}_{i=0}^{\chi_{\rm max} - 1}$ in each bond or to variationally find fixed-size tensors in $\mathfrak{B}$ by maximising the normalised Hilbert-Schmidt scalar product for $\mathfrak{A}$ and $\mathfrak{B}$ \cite{hubig-2017}. In both cases, the compression error can be quantified by the Frobenius norm $\|\mathfrak{A} - \mathfrak{B}\|_2$ (equivalent to the Hilbert-Schmidt norm and the Schatten 2-norm in our finite dimensional case). In the singular-value-truncation method, the upper bound is known, namely, $\|\mathfrak{A} - \mathfrak{B}\|_2^2 \leq 2 \sum_{\rm bonds} \sum_{i = \chi_{\rm max}}^{\chi-1} \lambda_i^2$ \cite{verstraete-2006}. However, in both cases the compression error can be calculated as $\|\mathfrak{A} - \mathfrak{B}\|_2^2 = {\rm tr}[\mathfrak{A}^2] + {\rm tr}[\mathfrak{B}^2] - 2 {\rm Re}({\rm tr}[\mathfrak{A}^{\dag}\mathfrak{B}])$. 

Construction of the noise-mitigation map ${\cal M}$ via iterative applications of Eq.~\eqref{middle-out-iteration} assumes that $i$th iteration map ${\cal M}^{(i)}$ is compressed down to bond dimension $\chi_{\rm max}$ if the actual bond dimension exceeds this value. Since the norm respects the triangle inequality, we upper bound the total error in the final compressed MPO $\mathfrak{M}_{\rm compr.}$ for ${\cal M}$ by
\begin{equation}
    \|\mathfrak{M}_{\rm compr.} - \mathfrak{M}_{\rm exact}\|_2 \leq  \sum_{i = 1}^{L} \|\mathfrak{M}_{\rm compr.}^{(i)} - \mathfrak{M}_{\rm exact}^{(i)}\|_2,
\end{equation}
where $L$ is the circuit depth.

Let $\mathfrak{r}$ be the PTM representation for the quasistate (Appendix \ref{appendix-rho}), $\mathfrak{o}$ be the PTM representation for the observable $O$ (Appendix \ref{appendix-H}). The noisy energy estimate is $\bar{O}_{\rm noisy} = \mathfrak{r}^{\dag} \mathfrak{o}$ and the noise mitigated value is $\bar{O}_{\rm n.m.} = \mathfrak{r}^{\dag} \mathfrak{M} \mathfrak{o}$. The compression error results in the energy estimate error $\Delta_{compr.}\bar{O}_{\rm n.m.}$ that can be bounded from above as follows:
\begin{eqnarray}
    \Delta_{\rm compr.}\bar{O}_{\rm n.m.} &=& |\mathfrak{r}^{\dag} (\mathfrak{M}_{\rm compr.} - \mathfrak{M}_{\rm exact}) \mathfrak{o}| \nonumber\\
    & \leq & |\mathfrak{r}| \cdot \|(\mathfrak{M}_{\rm compr.} - \mathfrak{M}_{\rm exact}) \| \cdot |\mathfrak{o}| \nonumber\\
    & \leq & |\mathfrak{r}| \cdot \|(\mathfrak{M}_{\rm compr.} - \mathfrak{M}_{\rm exact}) \|_2 \cdot |\mathfrak{o}| \nonumber\\
    & \leq & |\mathfrak{r}| \cdot |\mathfrak{o}| \cdot \sum_{i = 1}^{L} \|\mathfrak{M}_{\rm compr.}^{(i)} - \mathfrak{M}_{\rm exact}^{(i)}\|_2, \qquad \label{final-upper-bound-compression-error}
\end{eqnarray}
where $\|\bullet\| = \|\bullet\|_{\infty}$ is the conventional operator norm (the Schatten $\infty$-norm). The first inequality already gives not a very thight upper bound because for the $N$ qubit observable $O = \sum_{\boldsymbol{\alpha}} c_{\boldsymbol{\alpha}} \sigma_{\boldsymbol{\alpha}}$ we have $|\mathfrak{o}| = \sqrt{2^N \sum_{\boldsymbol{\alpha}} c_{\boldsymbol{\alpha}}}$, whereas $|\mathfrak{r}^{\dag}  \mathfrak{o}| \leq \sqrt{\sum_{\boldsymbol{\alpha}} c_{\boldsymbol{\alpha}}}$. Note that $|\mathfrak{r}|^2 = {\rm tr}[\varrho^2]$ is the quasistate purity parameter which continuously decreases with the increase of $L$ if the noise is unital. For example, if the noisy maps are two-qubit Pauli channels as in Sec.~\ref{section-two-qubit-pauli-channel} and different Pauli strings appear in $\varrho$ with the same frequency, then $|\mathfrak{r}| \approx \left(1- \frac{1}{16}\sum_{ij} \epsilon_{ij} \right)^{\# {\rm noisy~channels}}$. This behaviour partially compensates the growth of the norm $\|\mathfrak{M}_{\rm compr.} - \mathfrak{M}_{\rm exact} \|_2$. In fact, the operator $\mathfrak{M}_{\rm exact}$ expands the space of generalised Bloch vectors for $\varrho$ in exactly the opposite way and $\|\mathfrak{M}_{\rm exact} \|_2 \propto \left(1 + \frac{1}{16}\sum_{ij} \epsilon_{ij} \right)^{\# {\rm noisy~channels}}$ in Sec.~\ref{section-two-qubit-pauli-channel}. The final upper bound \eqref{final-upper-bound-compression-error} is usually too loose in practice, partially because of the drastic difference between the conventional operator norm $\|\bullet\|$ and the Frobenius norm $\|\bullet\|_{2}$ (the transition from the former one to the latter one was used in derivation of inequality \eqref{final-upper-bound-compression-error}) and partially by overestimating $|\mathfrak{r}^{\dag} {\mathfrak I} \mathfrak{o}|$ by $|\mathfrak{r}| \cdot |\mathfrak{o}|$. For example, the identity transformation ${\rm Id}$ for $N$ qubits in the PTM form is the $4^N \times 4^N$ identity matrix $\mathfrak{I}$ for which $\|\mathfrak{I}\| = 1$ whereas $\|\mathfrak{I}\|_{2} = 2^N$. A heuristic normalisation is typically used to get a reasonable error scaling. In our case, a division by $\|\mathfrak{M}_{\rm exact} \|_2$ enables us to get rid of  $|\mathfrak{r}|$ on one hand (as this division reproduces the scaling of $|\mathfrak{r}|$) and amend the overestimation of the operator norm on the other hand. We obtain a heuristic error estimate
\begin{equation}  \label{heuristic-error}
    \Delta_{\rm compr.}\bar{O}_{\rm n.m.} \sim \sqrt{\sum_{\boldsymbol{\alpha}} c_{\boldsymbol{\alpha}}} \cdot \sum_{i = 1}^{L} \frac{\|\mathfrak{M}_{\rm compr.}^{(i)} - \mathfrak{M}_{\rm exact}^{(i)}\|_2}{\|\mathfrak{M}_{\rm exact}^{(i)} \|_2}.
\end{equation}

\section{Distribution of MPO singular values} \label{appendix-sv-distribution}

\begin{figure}
    \centering
    \includegraphics[width = 8.5cm]{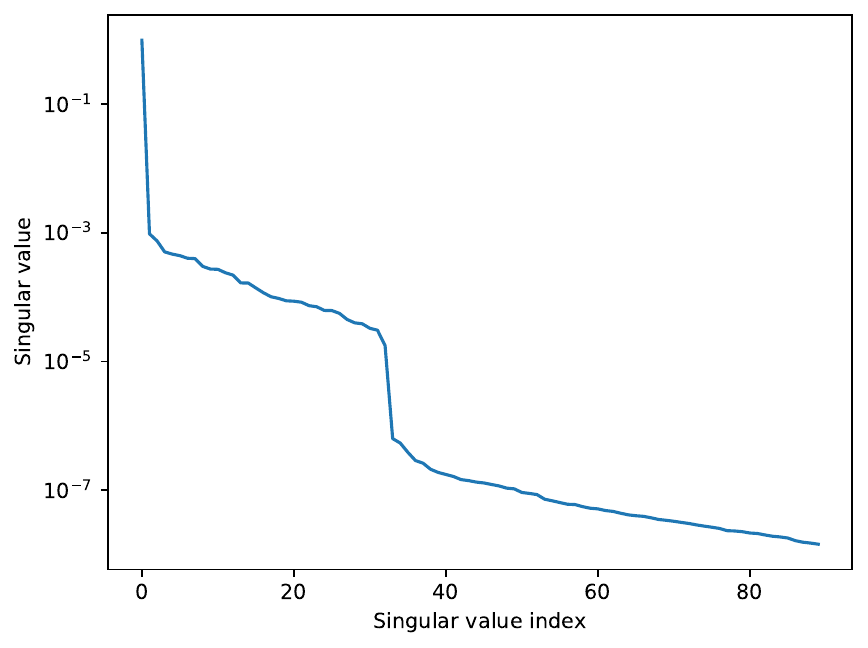} 
    \caption{Typical singular values in the central link for the MPO ${\cal M}$ (arranged in the decreasing order).}
    \label{figure-sv-distribution}
\end{figure}

Fig.~\ref{figure-sv-distribution} depicts typical singular values in the central link for the MPO ${\cal M}$ (arranged in the decreasing order). There is one leading singular value ($\sim 1$) and a plateau of singular values that are of the first order in the noise intensity ($\sim \epsilon$). Singular values exhibit transitions to higher orders in the noise intensity ($\sim \epsilon^2$).

\section{Stabiliser circuits} \label{appendix-stabilizer}

The Gottesman--Knill theorem insures a classically efficient simulation of stabiliser quantum circuits consisting of the Clifford gates \cite{gottesman-1998,aaronson-2004,anders-2006,gidney-2021}. Therefore, the stabiliser circuits serve as a natural testbed for studying scalability of quantum informational protocols (including the noise mitigation of the Clifford errors). The Clifford noise is a bistochastic quantum channel whose Kraus operators are proportional to the Clifford unitaries, which makes it possible to efficiently simulate the effect of the Clifford noise via probabilistic classical computation. Our numerical experiments are aimed at mitigating such a noise in exactly the same way as described in the proposed noise mitigation strategy (Sec.~\ref{section-middle-out} in the main text). 

\begin{figure*}
    \centering
    \includegraphics[width = 16cm]{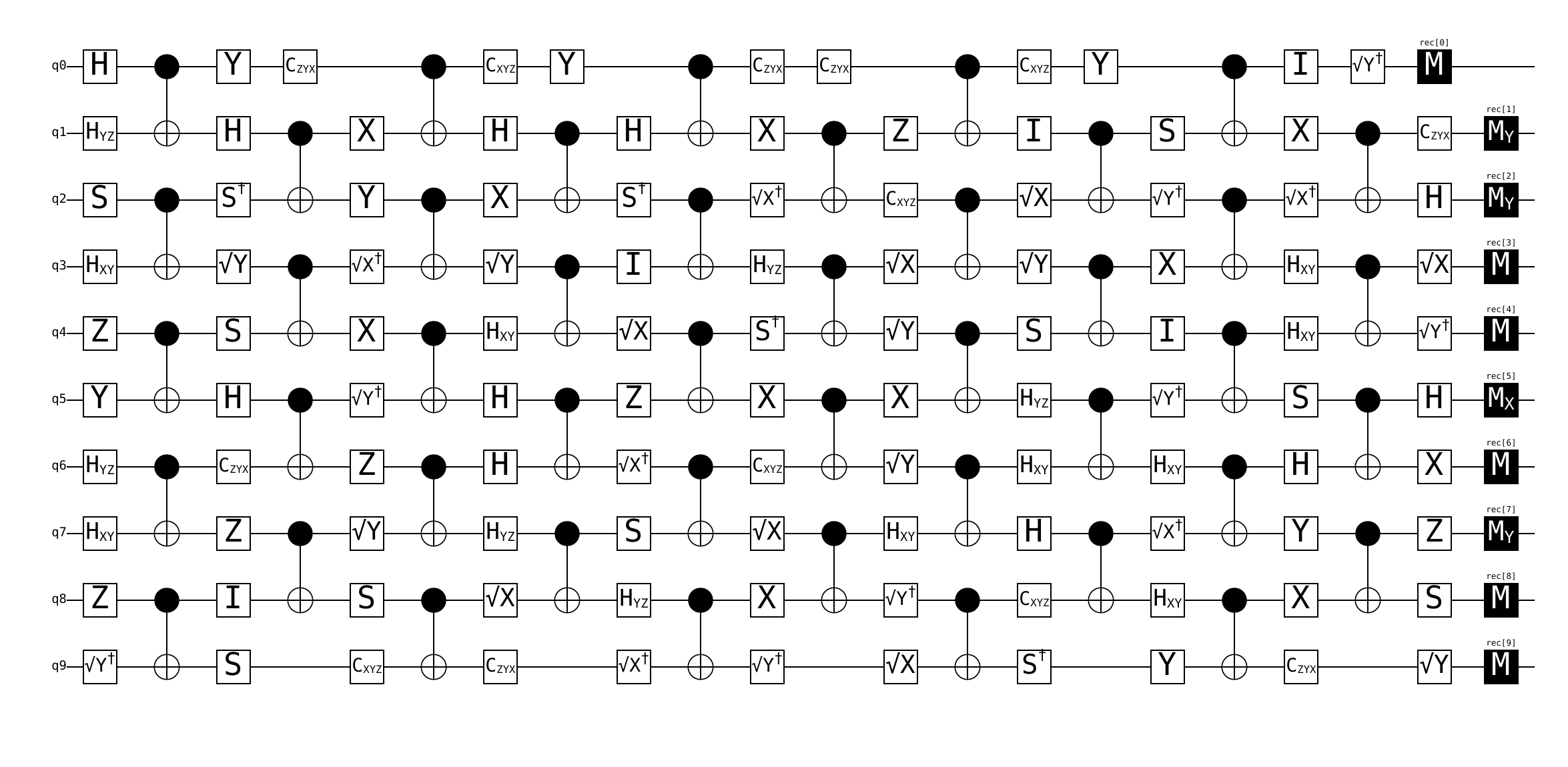} 
    \caption{An example of the stabiliser circuit with brick-wall-arranged layers of concurrent {\sc cnot} gates interleaved with layers of randomly chosen single-qubit Clifford gates. (Figure is created with the help of {\sc Stim} package \cite{gidney-2021}.)}
    \label{figure-stabilizer}
\end{figure*}

As the stabiliser circuit, we consider repeated brick-wall-arranged layers of concurrent {\sc cnot} gates (one starting at even and one at odd locations in the linear qubit register) interleaved with layers of randomly chosen single-qubit Clifford gates (Fig.~\ref{figure-stabilizer}). This enables the fastest propagation of correlations in the circuit. For a fixed number of qubits $N$ and the circuit depth $L$, the noiseless circuit prepares a generally correlated pure state vector $\ket{\psi}$, which is stabilised by all $N$ generating operators $\{g_i\}_{i=0}^{N-1}$ of the stabiliser group, i.e., $g_i \ket{\psi} = \ket{\psi}$ for all $i=0,\ldots,N$. To get a noisy version of the stabiliser circuit, each {\sc cnot} layer is followed by a sparse Pauli-Lindblad noise that just scales each of the stabilisers.

Once the noisy circuit prepares the density operator $\varrho$, the qubits are projectively measured in the eigenbasis of one of the Pauli operators. To get a non-zero estimation, the measurement basis for the whole circuit should be aligned with the eigenbasis of at least one generator $g_i$. Other bases can be optionally added for the sake of informational completeness but in this numerical experiment their probability can be made negligibly small as they do not usually contribute to the estimation of observable. Note that $({\cal N}^{-1})^{\dag}(g_i)$ and ${\cal U}^{\dag}(g_i)$ are both diagonal in the eigenbasis of $g_i$ for any Clifford noise ${\cal N}$ and any Clifford unitary operation ${\cal U}$. To sum up, the noisy circuit is measured in a specific local basis, with $S$ shots being collected. A fast simulation of measurement outcomes is possible with the help of the {\sc Stim} package \cite{gidney-2021}.

Without the noise mitigation, the estimation of a stabilizer $O_L$ gradually decreases from $1$ to $0$ with the increase of the circuit depth $L$ due to the noise accumulation. The estimated standard deviation $\Delta\bar{O_L}_{\rm noisy}$ is about $\frac{1}{\sqrt{S}}$ and does not depend on $L$. Application of the noise mitigation map amends the observable estimation and returns it back to the vicinity of the value $1$; however, the standard deviation $\Delta\bar{O_L}_{\rm n.m.}$ for the noise mitigated value increases. In the numerical experiment, the measurement overhead is the ratio $\gamma = \Delta\bar{O_L}_{\rm n.m.} / \Delta\bar{O_L}_{\rm noisy}$. The square $\gamma^2$ quantifies the scaling factor for the number of shots needed to reduce $\Delta\bar{O_L}_{\rm n.m.}$ down to $\Delta\bar{O_L}_{\rm noisy}$.

\section{Perturbation theory for the local Pauli noise in stabiliser circuits} \label{appendix-noise-in-stabilizer-circuits}

Consider a noisy stabiliser circuit, where each noisy layer ${\cal N}$ can be decomposed into a concatenation of $k$-local Pauli channels ${\cal N}^{[m_1,\ldots,m_k]}$ affecting $k$ qubits only (the qubits $m_1,\ldots,m_k$ do not have to be adjacent). A prominent example of $2$-local Pauli noise is the sparse Pauli-Lindblad noise model with the nearest-neighbour crosstalk (Sec.~\ref{appendix-pauli-lindblad-noise-model}). The leading term in each map ${\cal N}^{[m_1,\ldots,m_k]}$ is the identity transformation, so ${\cal N}^{[m_1,\ldots,m_k]} = {\rm Id} + \epsilon \Lambda^{[m_1,\ldots,m_k]}$, where $\epsilon$ is the noise intensity and $\Lambda$ is a $k$-local trace-nullifying map adopting a diagonal sum representation
$\Lambda^{[m_1,\ldots,m_k]}(\bullet) = \sum_{\alpha_{m_1},\ldots,\alpha_{m_k} = 0}^3 q_{\alpha_{m_1},\ldots,\alpha_{m_k}} \sigma_{\alpha_{m_1},\ldots,\alpha_{m_k}} \bullet \sigma_{\alpha_{m_1},\ldots,\alpha_{m_k}}$
with at most $4^k$ Kraus-like operators, each being proportional to a weight-$k$ Pauli string $\sigma_{\alpha_{m_1},\ldots,\alpha_{m_k}} \equiv I \otimes \cdots \otimes I \otimes \sigma_{\alpha_{m_1}} \otimes I \otimes \cdots \otimes I \otimes \sigma_{\alpha_{m_k}} \otimes I \otimes \cdots \otimes I$. 

\begin{figure}
    \centering
    \includegraphics[width = 8.5cm]{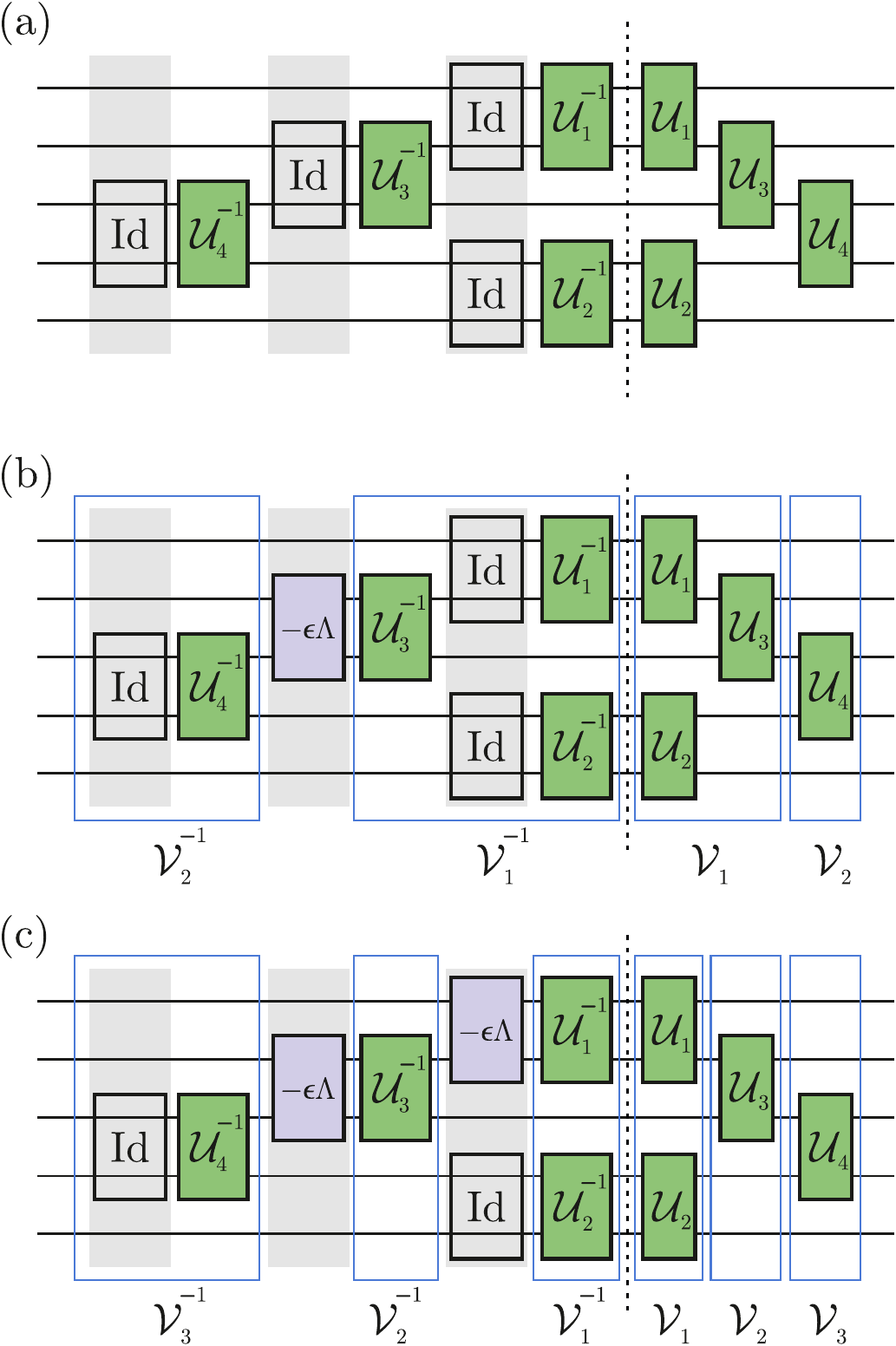}
    \caption{Perturbation theory for the noise-mitigation map with respect to the noise strength: (a) the only zero-order contribution, (b) a first-order contribution, (c) a second-order contribution.}
    \label{figure-perturbation}
\end{figure}

Let us develop a perturbation theory for the noise mitigation map ${\cal M}$ with respect to the noise intensity $\epsilon$. The zero-order contribution in ${\cal M}$ is simply the identity map ${\rm Id}$. To find the first-order contribution, we need to consider a particular map $\Lambda^{[m_1,\ldots,m_k]}$ and fix all other noisy maps to be the identity transformations, then sum over choices for $\Lambda^{[m_1,\ldots,m_k]}$ (see Fig.~\ref{figure-perturbation}). Suppose $\Lambda^{[m_1,\ldots,m_k]}$ intervenes in between unitary subcircuit operations ${\cal V}_1(\bullet) = V_1 \bullet V_1^{\dag}$ and ${\cal V}_2 = V_2 \bullet V_2^{\dag}$, then the corresponding first-order contribution to ${\cal M}$ is
\begin{equation} \label{map-1st-order-component}
    -\epsilon {\cal V}_2 {\cal V}_1 {\cal V}_1^{-1} \Lambda^{[m_1,\ldots,m_k]} {\cal V}_2^{-1}    = -\epsilon {\cal V}_2 \Lambda^{[m_1,\ldots,m_k]} {\cal V}_2^{-1}
\end{equation}
and has only $4^k$ Kraus-like operators proportional to $V_2 \sigma_{\alpha_{m_1},\ldots,\alpha_{m_k}} V_2^{\dag}$. Since $V_2$ is a stabiliser circuit itself, $V_2 \sigma_{\alpha_{m_1},\ldots,\alpha_{m_k}} V_2^{\dag}$ is a Pauli string, i.e., a factorised operator. Each map \eqref{map-1st-order-component} is a sum of $4^k$ factorised maps and can be exactly reproduced by an MPO tensor network with the bond dimension at most $4^k$. Therefore, all the first order contributions in the noise mitigation map are presented in a single MPO tensor network whose the bond dimension is at most $4^k$ times the total number of noisy $k$-local Pauli maps present throughout all noisy levels. For the sparse Pauli-Lindblad noise model with the nearest-neighbour crosstalk (Sec.~\ref{appendix-pauli-lindblad-noise-model}), $k=2$ and the total number of noisy $k$-local Pauli maps equals $(N-1)L$, where $N$ is the number of qubits and $L$ is the circuit depth. This means that if we choose the bond dimension $\chi_{\rm max} > 16(N-1)L$ while constructing the MPO for the noise mitigation map ${\cal M}$, then ${\cal M}$ captures all first-order noise contributions and the possible truncation error is at most of the second order $\epsilon^2$ in the noise intensity.

Similar arguments are applicable in all orders of the perturbation theory. For example, to find the second-order contribution to ${\cal M}$, we need to consider two particular maps $\Lambda_1^{[m_1^{(1)},\ldots,m_k^{(1)}]}$ and $\Lambda_2^{[m_1^{(2)},\ldots,m_k^{(2)}]}$, and fix all other noisy maps to be the identity transformations, then sum over choices for $\Lambda_1^{[m_1^{(1)},\ldots,m_k^{(1)}]}$ and $\Lambda_2^{[m_1^{(2)},\ldots,m_k^{(2)}]}$ (see Fig.~\ref{figure-perturbation}). Suppose $\Lambda_1^{[m_1^{(1)},\ldots,m_k^{(1)}]}$ and $\Lambda_2^{[m_1^{(2)},\ldots,m_k^{(2)}]}$ intervene in between three unitary subcircuit operations: ${\cal V}_1(\bullet) = V_1 \bullet V_1^{\dag}$, ${\cal V}_2 = V_2 \bullet V_2^{\dag}$, and ${\cal V}_3 = V_3 \bullet V_3^{\dag}$. Then the corresponding second-order contribution to ${\cal M}$ is
\begin{eqnarray} \label{map-2nd-order-component}
    && \!\!\!\!\!\!\!\! \epsilon^2 {\cal V}_3 {\cal V}_2 {\cal V}_1 {\cal V}_1^{-1} \Lambda_1^{[m_1^{(1)},\ldots,m_k^{(1)}]} {\cal V}_2^{-1} \Lambda_2^{[m_1^{(2)},\ldots,m_k^{(2)}]} {\cal V}_3^{-1} \nonumber\\
    && \!\!\!\!\!\!\!\! = \epsilon^2 \left( {\cal V}_3 {\cal V}_2 \Lambda_1^{[m_1^{(1)},\ldots,m_k^{(1)}]} {\cal V}_2^{-1} {\cal V}_3^{-1} \right) \left( {\cal V}_3 \Lambda_2^{[m_1^{(2)},\ldots,m_k^{(2)}]} {\cal V}_3^{-1} \right) \nonumber\\
\end{eqnarray}
and has only $4^{2k}$ Kraus-like operators proportional to $(V_3 V_2 \sigma_{\alpha_{m_1^{(1)}},\ldots,\alpha_{m_k^{(1)}}} V_2^{\dag} V_3^{\dag} ) (V_3 \sigma_{\alpha_{m_1^{(2)}},\ldots,\alpha_{m_k^{(2)}}} V_3^{\dag})$. Since $V_3 V_2$ and $V_3$ are stabiliser circuits themselves, $V_3 V_2 \sigma_{\alpha_{m_1^{(1)}},\ldots,\alpha_{m_k^{(1)}}} V_2^{\dag} V_3^{\dag}$ and $V_3 \sigma_{\alpha_{m_1^{(2)}},\ldots,\alpha_{m_k^{(2)}}} V_3^{\dag}$ are Pauli strings, and their product is again a Pauli string, i.e., a factorised operator. Each map \eqref{map-2nd-order-component} is a sum of $4^{2k}$ factorised maps and can be exactly reproduced by an MPO tensor network with the bond dimension at most $4^{2k}$. Therefore, all the second order contributions in the noise mitigation map are presented in a single MPO tensor network whose the bond dimension is at most $4^{2k}$ times the binomial coefficient $\binom{\#_{\rm noisy}}{2}$, where $\#_{\rm noisy}$ is the total number of noisy $k$-local Pauli maps present throughout all noisy levels. For the sparse Pauli-Lindblad noise model with the nearest-neighbour crosstalk (Sec.~\ref{appendix-pauli-lindblad-noise-model}), $k=2$ and $\#_{\rm noisy} = (N-1)L$, where $N$ is the number of qubits and $L$ is the circuit depth. This means that if we exceed the threshold bond dimension ($\chi_{\rm max} > 16(N-1)L + 128(N-1)L[(N-1)L-1] \approx 128 N^2 L^2$) while constructing the MPO for the noise mitigation map ${\cal M}$, then ${\cal M}$ captures all first-and second-order noise contributions and the possible truncation error is at most of the third order $\epsilon^3$ in the noise intensity.

Using a property of binomial coefficients (based on \cite{sum-binomial}),
\begin{equation*}
    \sum_{i=0}^l \binom{\#}{i} x^i \leq \frac{\# - l + 1}{\# - l+1 - \frac{l}{x}} \binom{\#}{l} x^l,
\end{equation*}
and Stirling's approximation, we conclude that the possible truncation error cannot exceed the order of $\epsilon^{l+1}$ if the bond dimension $\chi_{\rm max} > 4^{lk} \#_{\rm noisy}^l / l!$. 

Conversely, for a given maximum bond dimension $\chi_{\rm max}$ used in compression of the noise mitigation map, the compression error cannot exceed of the order $\epsilon^{\lceil l \rceil}$, where the approximate value of $l$ is found by exploiting Stirling's approximation and an iterative method (up to the second iteration): 
\begin{equation}
    l \approx l_2 = \frac{\ln \chi + \frac{1}{2}\ln(2\pi l_1)}{\ln(4^k \#_{\rm noisy}) + 1 - \ln l_1}, \quad l_1 = \frac{\ln \chi}{\ln (4^k \#_{\rm noisy})}.
\end{equation}
For the sparse Pauli-Lindblad noise model with the nearest-neighbour crosstalk (Sec.~\ref{appendix-pauli-lindblad-noise-model}), scaling of the compression error is roughly $\epsilon^{\log_{16NL}(\chi)}$. Any desired power of $\epsilon$ is achievable with the bond dimension $\chi$ polynomially scaling in the number of circuit gates ($\sim NL$).

\section{Measurement overhead} \label{appendix-measurement-overhead}

\begin{figure}
    \centering
    \includegraphics[width = 8.5cm]{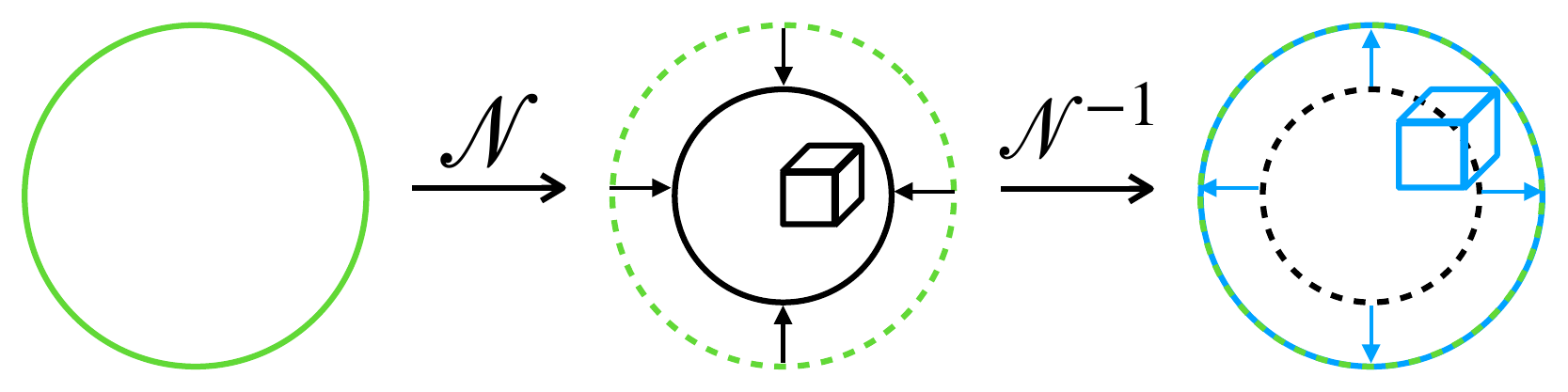} 
    \caption{Pictorial representation of the generalised Bloch ball transformation due to noise and noise inversion. The estimation error (black box in the middle) is amplified by the noise mitigation map.}
    \label{figure-state-expansion}
\end{figure}

In the proposed TEM strategy, the resulting estimation error $\Delta \bar{O}_{\rm n.m.}$ is greater than the noisy estimation $\Delta \bar{O}_{\rm noisy}$ due to the presence of inverse maps ${\cal N}^{-1}$ in  $\mathcal{M}$. These inverse maps expand the state space (in the generalised PTM representation for $N$-qubit states) and govern the mixed density operator $\varrho$ at the noisy circuit output to a pure state $\ket{\psi}\bra{\psi}$ that the corresponding noiseless circuit would produce. Fig.~\ref{figure-state-expansion} pictorially explains this effect at the level of states; however, the relation between $\Delta \bar{O}_{\rm n.m.}$ and $\Delta \bar{O}_{\rm noisy}$ depends not only on the noisy density operator $\varrho$ and the noisy circuit but also on the observable $O$. For example, if $O$ is close to the identity operator and the noise is unital, then $\Delta \bar{O}_{\rm n.m.} = \Delta \bar{O}_{\rm noisy}$ manifesting no measurement overhead. 

To make the last argument clearer and benchmark against the PEC measurement overhead, let us consider an example of the single-qubit depolarising noise ${\cal N}[\varrho] = (1-\epsilon) \varrho + \epsilon {\rm tr}[\varrho] \frac{1}{2} I$. The Kraus-like representation of the inverse map (which is neither completely positive nor positive if $\epsilon > 0$) reads
\begin{equation*}
    {\cal N}^{-1}(\varrho) \approx \left(1+\tfrac{3\epsilon}{4} \right) \rho -  \tfrac{\epsilon}{4}(X \varrho X + Y \varrho Y + Z \varrho Z),
\end{equation*}
where $(X,Y,Z) \equiv (\sigma_1,\sigma_2,\sigma_3)$. In the PEC, ${\cal N}^{-1}$ is simulated by sampling Pauli gates $I,X,Y,Z$ from the quasiprobability $(1+\tfrac{3\epsilon}{4},-\tfrac{\epsilon}{4},-\tfrac{\epsilon}{4},-\tfrac{\epsilon}{4})$, which implies sampling from the actual probability distribution $\gamma_{\rm PEC}^{-1} \times (1+\tfrac{3\epsilon}{4},\tfrac{\epsilon}{4},\tfrac{\epsilon}{4},\tfrac{\epsilon}{4})$ with the overhead $\gamma_{\rm PEC} \approx 1 + \tfrac{3\epsilon}{2}$. One can see two different contributions to $\gamma_{\rm PEC}$: one originates from the amplifying factor $1+\tfrac{3\varepsilon}{4}$ and the other one accounts for negativities in the quasiprobability ($3 \times \frac{\epsilon}{4}$). In the TEM, the map ${\cal N}^{-1}$ is applied as a mathematical map in the classical post-processing. To estimate the overhead in this case, we formally consider the evolution of an observable $O$ in the Heisenberg picture $(\mathcal{N}^{-1})^{\dag} = \mathcal{N}^{-1}$ (though the map ${\cal N}^{-1}$ is not completely positive). If $O$ is one of the Pauli operators, then
\begin{eqnarray*}
    && \mathcal{N}^{-1} [I] = I, \\ 
&& \mathcal{N}^{-1} [(X,Y,Z)] = \tfrac{1}{1-\epsilon} (X, Y, Z) \approx (1+\epsilon) \times (X, Y, Z).
\end{eqnarray*}
The measurement overhead in this case $\gamma_{\rm TEM} \leq 1+\epsilon < \gamma_{\rm PEC}$. If $O$ has same-order contributions from all Pauli operators, then we get the averaged measurement overhead
\begin{equation} \label{dep-1-gammas}
    \gamma_{\rm TEM} \approx \tfrac{1}{4} \times 1 + \tfrac{3}{4} \times (1+ \epsilon) = 1 + \tfrac{3\epsilon}{4} \approx \sqrt{\gamma_{\rm PEC}}.
\end{equation}
In the TEM, there is only one contribution to $\gamma_{\rm TEM}$ associated with the amplification factor.

The same line of reasoning is applicable to the 2-qubit depolarising noise \eqref{2-qubit-dep}. In this case, the inverse map \eqref{2-qubit-dep-inverse} takes the form
\begin{equation*}
    {\cal N}^{-1}(\varrho) \approx \left(1+\tfrac{15\epsilon}{16} \right) \rho -  \tfrac{\epsilon}{16} (I \otimes X \varrho I \otimes X + \ldots + Z \otimes Z \varrho Z \otimes Z),
\end{equation*}
the quasiprobability distribution is $(1+\tfrac{15\epsilon}{16},-\tfrac{\epsilon}{16},\ldots,-\tfrac{\epsilon}{16})$ and $\gamma_{\rm PEC} \approx 1+\frac{15\epsilon}{8}$. On the other hand, in the TEM
\begin{eqnarray*}
    && \mathcal{N}^{-1} [I \otimes I] = I \otimes I, \\ 
&& \mathcal{N}^{-1} [(I \!\! \otimes \!\! X,\ldots,Z \!\! \otimes\! \! Z)] \approx (1+\epsilon) \times (I \!\! \otimes \!\! X,\ldots,Z \!\! \otimes \!\! Z).
\end{eqnarray*}
If the noise affects 2 of $N$ qubits, then the observable's Pauli substrings (affecting those 2 qubits) are relevant for the measurement overhead analysis. If most of the substrings are identity operators (as it happens for a low-Pauli-weight observable $O$), then the measurement overhead is negligible (close to $1$). Otherwise, if all 16 substrings appear with roughly the same frequency and the same-order coefficients, then we get the averaged measurement overhead
\begin{equation} \label{dep-2-gammas}
    \gamma_{\rm TEM} \approx \tfrac{1}{16} \times 1 + \tfrac{15}{16} \times (1+ \epsilon) = 1 + \tfrac{15\epsilon}{16} \approx \sqrt{\gamma_{\rm PEC}}.
\end{equation}
If the circuit contains $\#_{\rm noisy}$ 2-qubit depolarising maps, then the measurement overheads for a typical observable are
\begin{equation} \label{gamma-TEM-2-dep-N}
    \gamma_{\rm PEC} \approx \left( 1 + \frac{15\epsilon}{8} \right)^{\#_{\rm noisy}}, \quad \gamma_{\rm TEM} \approx \left( 1 + \frac{15\epsilon}{16} \right)^{\#_{\rm noisy}}.
\end{equation}

Eqs.~\eqref{dep-1-gammas} and \eqref{dep-2-gammas} [altogether with similar calculations for a general 2-qubit Pauli noise (Sec.~\ref{section-two-qubit-pauli-channel})] reflect a general square-root relation $\gamma_{\rm TEM} \approx \sqrt{\gamma_{\rm PEC}}$ for typical high-Pauli-weight observables under the Pauli noise (discussed in the main text in Sec.~\ref{section-middle-out}). The low-Pauli-weight observables enjoy even small measurement overhead, which makes our approach beneficial for estimating two-point correlators ($k$-local correlators, $k \ll N$) and chemical Hamiltonians (whose Pauli weight generally grows logarithmically in the number of qubits $N$ \cite{miller-2022}). 

Interestingly, the averaged measurement overhead in the TEM can be inferred from the very noise mitigation map ${\cal M}$. In Sections \ref{section-two-qubit-depolarizing} and \ref{section-two-qubit-pauli-channel}, we present singular values in the MPO link for the inverse of the 2-qubit depolarising noise and the 2-qubit Pauli noise, respectively. The largest singular value $\Sigma_{0}^{0}$ is an effective amplification factor associated with the identity transformation. Therefore, the measurement overhead in the TEM is readily estimated as the largest singular value in the MPO for ${\cal M}$ (regularised w.r.t. the singular value of the identity transformation). 

\section{Instability in the zero-noise extrapolation} \label{appendix-zne}

\begin{figure}[t]
    \centering
    \includegraphics[width = 0.5\textwidth]{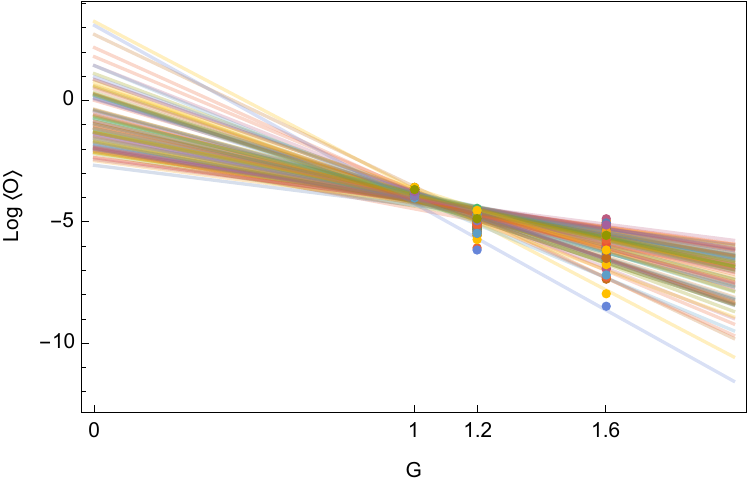}
    \caption{Instability in the exponential zero-noise extrapolation through noisy values estimated in a deep circuit and clustering around zero. Dots depict the bootstrapped noisy estimations for different values of the noise gain $G$.}
    \label{figure-zne-instability}
\end{figure}

In ZNE-PEA, the noise is amplified intentionally with various gain factors so as to extrapolate the noisy estimations to the zero noise gain and get the noise-mitigated observable estimation. Generally, different extrapolations can be used (e.g., linear and exponential) and the noise-mitigated estimation can be biased, meaning that the estimation can differ from the exact value even in the limit of infinitely many shots. For a Pauli string observable and a stabiliser circuit with the sparse Pauli-Lindblad noise the exponential extrapolation is known to be unbiased; however, it can still be very imprecise with a finite number of shots available. This happens if the noisy values cluster around the same value so that the difference between them becomes comparable with the shot noise. This is exactly the case for the observables and deep circuits presented in Sec.~\ref{section-results}. To demonstrate ZNE-PEA instability, we consider extrapolation through three noise-gain values $G=1.0, 1.2, 1.6$ used in the recent experiment~\cite{kim-2023}. For a stabiliser circuit with the overall measurement overhead $\gamma_{\rm PEC} > 2 \times 10^3$, the corresponding noisy estimates $\bar{O}_{\rm n.}(G)$ are below $\sim 0.02$ which makes the barely distinguishable with the budget of $3 \times 10^5$ shots per single noise gain value. To estimate the final extrapolation result and the error bar in this case, we follow the lines of Ref.~\cite{kim-2023} and resort to bootstrapping. The measurement outcomes for each noise gain are resampled 100 times and averaged. Each of 100 tuples $\big(\bar{O}_{\rm n.}(1.0),\bar{O}_{\rm n.}(1.2),\bar{O}_{\rm n.}(1.6)\big)$ is extrapolated exponentially to get a set $\{\bar{O}_{\rm n.}(0)\}_{i=1}^{100}$. Fig.~\ref{figure-zne-instability} shows the extrapolation instability in this case. The median of this set gives the final extrapolation result. Then we throw away $16\%$ largest and $16\%$ smallest values in the set $\{\bar{O}_{\rm n.}(0)\}_{i=1}^{100}$. The range between the largest and the smallest values of the remaining set define the error bar.

\section{Algorithm for the TEM}

To summarise, the TEM algorithm is as follows:
\begin{enumerate}
    \item Given an ideal circuit (to be implemented later via a noisy hardware), find MPO representation for each unitary-map layer ${\cal U}$ and its conjugation ${\cal U}^{-1}$.
    \item Characterise the noise ${\cal N}$ accompanying each unitary layer and find MPO representation for ${\cal N}^{-1}$.
    \item Construct the noise-inversion map ${\cal M}$ in the middle-out fashion by sequential applications of Eq.~\eqref{middle-out-iteration} in terms of tensor networks: multiply the MPOs and compress the result (down to the maximum bond dimension $\chi_{\rm max}$ or the bond dimension adjusts to a predefined compression precision). 
    \item Monitor the compression error for it to be below some desired value or take it into account when estimating the observable (see Appendix \ref{appendix-convergence}).
    \item Implement circuit in the noise-characterised quantum hardware and collect samples from the IC measurements.
    \item Contract the whole tensor network in Fig.~\ref{figure-circuit}(b), i.e., measurement outcomes, dual operators, the noise-mitigation map ${\cal M}$, and the observable operator. The contraction is a single noise-mitigated value $\bar{O}_{\rm n.m.}$.
    \item Estimate error $\Delta\bar{O}_{\rm n.m.}$ in the observable estimation via the standard statistical methods and incorporate the compression error if the latter is comparable or large than the former.
    
\end{enumerate}

\section{Convergence of TEM for increasing bond dimension}\label{appendix-convergence}

\begin{figure}[t]
    \centering
    \includegraphics[width = 0.9\textwidth]{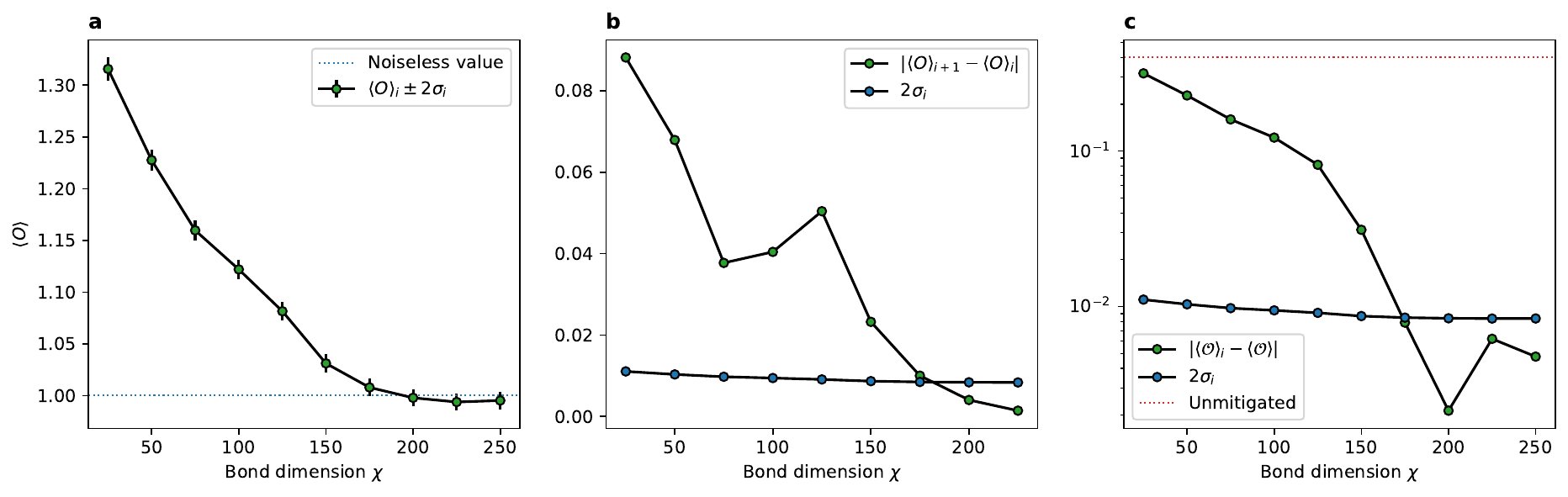}
    \caption{Convergence of TEM for increasing bond dimension.
    (a) For each bond dimension in the list $(25, 50, \ldots, 225, 250)$ we run the middle-out compression of the noise map and use the resulting map in TEM to produce the mitigated estimations $\langle O \rangle_i$, with standard error $\sigma_i$.
    (b) The green dots depict the change between the estimations obtained with two consecutive bond dimensions (i.e.~$\chi$ and $\chi + 25$).
    The blue dots show $2 \sigma_i$ for the estimation with bond dimension $\chi$.
    (c) Absolute error between each estimation and the exact, noiseless value (green) and standard error of each estimation (blue).
    The red dashed line shows the initial, unmitigated error.}
    \label{figure-convergence}
\end{figure}

As the bond dimension in each middle-out compression step is increased, the truncation error in the TEM map decreases, and so does the resulting error in different observables.
However, since different observables may be affected differently by the truncation error, it is advisable to assess the bond dimension required by the method on a case-by-case basis.
In short, the idea is to proceed in a similar fashion as in other tensor network approaches such as DMRG, and simply analyse how the mean value of the observable of interest changes with increasing bond dimension until convergence is reached.

To illustrate this, let us consider a particular example with a 100-qubit-wide and 20-layer-deep Clifford circuit like the ones used in the main text.
We add depolarising noise with $\epsilon = 0.001$.
In order to use a non-trivial observable, let us use the operator $O = U_L Z_{5} Z_{25} Z_{45} Z_{65} Z_{85} U_L^{\dagger}$, where $U_L$ is the Clifford circuit. $O$ is highly non-local, with $\sim 70 \%$ of its components being different from identity.

The noiseless expectation value of the observable is $\langle O \rangle_{\rm exact} = 1$.
Running the noisy circuit for $10^5$ measurement rounds yields the estimate $\langle O \rangle_{\rm noisy} = 0.6006$ with standard error $\sigma_{\rm noisy} = 0.0025$.
Next, we run TEM for several bond dimensions between 25 and 250 in increments of 25 and produce a set of estimated means $\langle O \rangle_i$ and estimated standard errors $\sigma_i$, where $i$ stands for the index in the list of bond dimensions used.
These are depicted in Fig.~\ref{figure-convergence}a, where it can be appreciated that the mitigated value quickly approaches the exact one.

In real applications, one may not know the exact value beforehand, so convergence must be assessed in some other way.
Now, since any estimate we produce will have some standard error (due to the finite sampling from the quantum computer), which can be estimated, we may consider TEM to have converged if subsequent increases in the bond dimension only produce changes of that order or smaller.
In Fig.~\ref{figure-convergence}b, we plot the difference between consecutive estimations of the mean for sufficiently large increments of bond dimension, and compare them against their standard errors.
As it can be seen, the differences in estimations with bond dimensions 225 and 200, as well as with 250 and 225, are smaller than their corresponding $2 \sigma_i$ statistical fluctuations, so one can consider the algorithm to have converged.

To verify this convergence criterion, in Fig.~\ref{figure-convergence}c we plot the error of each estimation with respect to the exact mean value.
The results confirm that increasing the bond dimension beyond 200 does not result in any benefits, as the errors are below the $2 \sigma_i$ error levels of the estimations.
Interestingly, however, this figure also reveals that the convergence of TEM is exponential in the bond dimension in the regime in which truncation errors dominate, that is, for bond dimension below 200 in this example.

\section{Parameters of the generated sparse Pauli-Lindblad noise model} \label{appendix-rates}

In the sparse Pauli-Lindblad models for the example of 10-qubit Trotter dynamics in Sec.~\ref{section-results}, we use the dimensionless rate coefficients $\lambda_{\boldsymbol{\alpha}}$ listed in Tables \ref{table-SPL-1} and \ref{table-SPL-2}. The jump operator $\sigma_{\boldsymbol{\alpha}}$ is associated with the noise map $\exp[\lambda_{\boldsymbol{\alpha}}(\sigma_{\boldsymbol{\alpha}} \bullet \sigma_{\boldsymbol{\alpha}} - \bullet)]$. The rates for the jump operators acting non-trivially on a single qubit are sampled from the normal distribution with the mean 0.001 and the standard deviation 0.0009. The rates for the jump operators acting non-trivially on two neighbour qubits are sampled from the normal distribution with the mean 0.0004 and the standard deviation 0.0006. Negative rates are reset to zero.

In the analysis of stabliser circuits, we follow the same parameter generation approach but with different parameters. The rates for the jump operators acting non-trivially on a single qubit are sampled from the normal distribution with the mean $6.67\times 10^{-5}$ and the standard deviation $6.67\times 10^{-5}$. The rates for the jump operators acting non-trivially on two neighbour qubits are sampled from the normal distribution with the mean $2.22\times 10^{-5}$ and the standard deviation $2.22\times 10^{-5}$. Negative rates are reset to zero. We omit the full tables with 1191 rate coefficients here. They are available upon request. The resulting effective noise per layer per qubit $\bar{\gamma}_{\rm PEC}-1 = 9.1 \times 10^{-4}$.

\section{Details on the numerical simulations}
\label{app:numerical_details}
The code for producing the results shown in Sec.~\ref{section-results} has been implemented using \textsc{Python}. For simulating stabiliser circuits, we used the \textsc{Stim} package \cite{gidney-2021}. For tensor network contractions and compression the library \textsc{quimb} \cite{quimb} was used. In particular, the compression of the bonds was performed using the randomized SVD (\texttt{rsvd()}) algorithm provided in the latter. As backend for the tensor algebra operations, \textsc{numpy} with Intel \textsc{MKL BLAS} support was used. To perform the TEM of the 100-qubit, 100-layer circuit described in Fig.~\ref{figure-stabilizer-results}, a total wall time of 63 hours was employed on an Azure \texttt{Standard\_HC44rs} machine (44 Intel Xeon Platinum 8168 vCPUs, year 2017). The high-water mark for the memory consumption is around 100 GB. It is worth noticing that no custom optimisation or parallelisation of the code has been performed outside of the built-in multi-threaded execution of the BLAS routines in MKL. We expect significant reduction in the wall  dtime from optimised tensor algebra backends and from using specialised acceleration hardware such as GPUs or TPUs, as well as distributed computation.

\begin{table}
\caption{\label{table-SPL-1} Parameters of the sparse Pauli-Lindblad model for the noise layer 1 in 10-qubit Trotter dynamics}
\begin{ruledtabular}
\begin{tabular}{c|c||c|c||c|c||c|c}
  Jump  & Decoherence  & Jump & Decoherence & Jump & Decoherence & Jump & Decoherence \\
  operator  & rate & operator & rate & operator & rate & operator & rate \\
  \hline
  $X_1$ &  0.0007188384538  & $Y_{10}$ & 0.0010735326374 & $Z_3 Z_4$ & 0.0 & $X_7 X_8$ & 0.0006800975765 \\
  $Y_1$ &  0.0026219497414  & $Z_{10}$ & 0.0009854359356 & $X_4 X_5$ & 0.0 & $X_7 Y_8$ & 0.0011992236472 \\
  $Z_1$ &  0.0015496680743  & $X_1 X_2$ & 0.0004171448785 & $X_4 Y_5$ & 0.0004145140852 & $X_7 Z_8$ & 0.0010657734054 \\
  $X_2$ &  0.0013697548210  & $X_1 Y_2$ & 0.0004000124513 & $X_4 Z_5$ & 0.0 & $Y_7 X_8$ & 0.0005693122825 \\
  $Y_2$ &  0.0015972844218  & $X_1 Z_2$ & 0.0 & $Y_4 X_5$ & 0.0002399826614 & $Y_7 Y_8$ & 0.0000110333258 \\
  $Z_2$ &  0.0026786081192  & $Y_1 X_2$ & 0.0013544305447 & $Y_4 Y_5$ & 0.0014391631248 & $Y_7 Z_8$ & 0.0005851477777 \\
  $X_3$ &  0.0009840379484  & $Y_1 Y_2$ & 0.0006013215169 & $Y_4 Z_5$ & 0.0 & $Z_7 X_8$ & 0.0008180286985 \\
  $Y_3$ &  0.0015345232914  & $Y_1 Z_2$ & 0.0001360847576 & $Z_4 X_5$ & 0.0 & $Z_7 Y_8$ & 0.0 \\
  $Z_3$ &  0.0020274482252  & $Z_1 X_2$ & 0.0002374506778 & $Z_4 Y_5$ & 0.0 & $Z_7 Z_8$ & 0.0002920344835 \\
  $X_4$ &  0.0010863884090  & $Z_1 Y_2$ & 0.0 & $Z_4 Z_5$ & 0.0011609124234 & $X_8 X_9$ & 0.0003609533666 \\
  $Y_4$ &  0.0001065144408  & $Z_1 Z_2$ & 0.0 & $X_5 X_6$ & 0.0 & $X_8 Y_9$ & 0.0006946695859 \\
  $Z_4$ &  0.0017631475265  & $X_2 X_3$ & 0.0 & $X_5 Y_6$ & 0.0020571240593 & $X_8 Z_9$ & 0.0011090370933 \\
  $X_5$ &  0.0  & $X_2 Y_3$ & 0.0014591789675 & $X_5 Z_6$ &  0.0 & $Y_8 X_9$ & 0.0004346218145 \\
  $Y_5$ &  0.0  & $X_2 Z_3$ & 0.0001143396436 & $Y_5 X_6$ & 0.0011758015780 & $Y_8 Y_9$ & 0.0 \\
  $Z_5$ &  0.0015495526495  & $Y_2 X_3$ & 0.0003231157903 & $Y_5 Y_6$ & 0.0003494956822 & $Y_8 Z_9$ & 0.0001599411535 \\
  $X_6$ &  0.0  & $Y_2 Y_3$ & 0.0003047123740 & $Y_5 Z_6$ & 0.0005244052712 & $Z_8 X_9$ & 0.0006980513828 \\
  $Y_6$ &  0.0023902359479  & $Y_2 Z_3$ & 0.0009723767925 & $Z_5 X_6$ & 0.0003472416886 & $Z_8 Y_9$ & 0.0 \\
  $Z_6$ &  0.0010958132238  & $Z_2 X_3$ & 0.0009163673808 & $Z_5 Y_6$ & 0.0002517186965 & $Z_8 Z_9$ & 0.0013271429858 \\
  $X_7$ &  0.0020257855184  & $Z_2 Y_3$ & 0.0014617357731 & $Z_5 Z_6$ & 0.0008080793202 & $X_9 X_{10}$ & 0.0009791413973 \\
  $Y_7$ &  0.0021400855405  & $Z_2 Z_3$ & 0.0001052694463 & $X_6 X_7$ & 0.0004431773843 & $X_9 Y_{10}$ & 0.0011589174283 \\
  $Z_7$ &  0.0005788805915  & $X_3 X_4$ & 0.0007171705339 & $X_6 Y_7$ & 0.0000341754743 & $X_9 Z_{10}$ & 0.0001831596684 \\
  $X_8$ &  0.0019446767565  & $X_3 Y_4$ & 0.0004325833089 & $X_6 Z_7$ & 0.0002432753686 & $Y_9 X_{10}$ & 0.0003006216905 \\
  $Y_8$ &  0.0014794883469  & $X_3 Z_4$ & 0.0017211740069 & $Y_6 X_7$ & 0.0009004397971 & $Y_9 Y_{10}$ & 0.0004491016202 \\
  $Z_8$ &  0.0004834382590  & $Y_3 X_4$ & 0.0010317935262 & $Y_6 Y_7$ & 0.0 & $Y_9 Z_{10}$ & 0.0010136247483 \\
  $X_9$ &  0.0014522106481  & $Y_3 Y_4$ & 0.0007849610247 & $Y_6 Z_7$ & 0.0008233462545 & $Z_9 X_{10}$ & 0.0010611984571 \\
  $Y_9$ &  0.0009070535380  & $Y_3 Z_4$ & 0.0009968302758 & $Z_6 X_7$ & 0.0007495748739 & $Z_9 Y_{10}$ & 0.0 \\
  $Z_9$ &  0.0023264494057  & $Z_3 X_4$ & 0.0 & $Z_6 Y_7$ & 0.0007152515147 & $Z_9 Z_{10}$ & 0.0012903240539 \\
  $X_{10}$ &  0.0006858528793  & $Z_3 Y_4$ & 0.0002774239964 & $Z_6 Z_7$ & 0.0012941373561 &  &  \\
\end{tabular}
\end{ruledtabular}
\end{table}

\begin{table}
\caption{\label{table-SPL-2}Parameters of the sparse Pauli-Lindblad model for the noise layer 2 in 10-qubit Trotter dynamics}
\begin{ruledtabular}
\begin{tabular}{c|c||c|c||c|c||c|c}
  Jump  & Decoherence  & Jump & Decoherence & Jump & Decoherence & Jump & Decoherence \\
  operator  & rate & operator & rate & operator & rate & operator & rate \\
  \hline
  $X_1$ &  0.0              & $Y_{10}$  & 0.0017389696194 & $Z_3 Z_4$ & 0.0017425011303 & $X_7 X_8$ & 0.0007253097676 \\
  $Y_1$ &  0.0014693600715  & $Z_{10}$  & 0.0026542233490 & $X_4 X_5$ & 0.0007077367641 & $X_7 Y_8$ & 0.0 \\
  $Z_1$ &  0.0026627276453  & $X_1 X_2$ & 0.0004237835913 & $X_4 Y_5$ & 0.0002408123508 & $X_7 Z_8$ & 0.0004465530507 \\
  $X_2$ &  0.0009620591518  & $X_1 Y_2$ & 0.0003630373210 & $X_4 Z_5$ & 0.0006666894811 & $Y_7 X_8$ & 0.0006274975328 \\
  $Y_2$ &  0.0010564395207  & $X_1 Z_2$ & 0.0002848231897 & $Y_4 X_5$ & 0.0000454529163 & $Y_7 Y_8$ & 0.0 \\
  $Z_2$ &  0.0009723465495  & $Y_1 X_2$ & 0.0             & $Y_4 Y_5$ & 0.0004523831237 & $Y_7 Z_8$ & 0.0006127668507 \\
  $X_3$ &  0.0013388355933  & $Y_1 Y_2$ & 0.0             & $Y_4 Z_5$ & 0.0003603372457 & $Z_7 X_8$ & 0.0012447982973 \\
  $Y_3$ &  0.0010427760033  & $Y_1 Z_2$ & 0.0005966276473 & $Z_4 X_5$ & 0.0             & $Z_7 Y_8$ & 0.0006178629978 \\
  $Z_3$ &  0.0              & $Z_1 X_2$ & 0.0             & $Z_4 Y_5$ & 0.0002380569629 & $Z_7 Z_8$ & 0.0006609217767 \\
  $X_4$ &  0.0023105048745  & $Z_1 Y_2$ & 0.0007309942553 & $Z_4 Z_5$ & 0.0004160620513 & $X_8 X_9$ & 0.0002185124723 \\
  $Y_4$ &  0.0014919297884  & $Z_1 Z_2$ & 0.0000015157442 & $X_5 X_6$ & 0.0             & $X_8 Y_9$ & 0.0005693242109 \\
  $Z_4$ &  0.0002082390859  & $X_2 X_3$ & 0.0015575541386 & $X_5 Y_6$ & 0.0011914765728 & $X_8 Z_9$ & 0.0007241280299 \\
  $X_5$ &  0.0009261012701  & $X_2 Y_3$ & 0.0             & $X_5 Z_6$ & 0.0008437638617 & $Y_8 X_9$ & 0.0007282630750 \\
  $Y_5$ &  0.0007442025724  & $X_2 Z_3$ & 0.0005837913797 & $Y_5 X_6$ & 0.0001668267636 & $Y_8 Y_9$ & 0.0010641609267 \\
  $Z_5$ &  0.0023443828473  & $Y_2 X_3$ & 0.0001372174639 & $Y_5 Y_6$ & 0.0009173528423 & $Y_8 Z_9$ & 0.0001106725915 \\
  $X_6$ &  0.0              & $Y_2 Y_3$ & 0.0001625124553 & $Y_5 Z_6$ & 0.0             & $Z_8 X_9$ & 0.0 \\
  $Y_6$ &  0.0007993187098  & $Y_2 Z_3$ & 0.0009019187150 & $Z_5 X_6$ & 0.0010055930809 & $Z_8 Y_9$ & 0.0020606323308 \\
  $Z_6$ &  0.0017095711268  & $Z_2 X_3$ & 0.0             & $Z_5 Y_6$ & 0.0000652947879 & $Z_8 Z_9$ & 0.0003172806200 \\
  $X_7$ &  0.0006089087327  & $Z_2 Y_3$ & 0.0002304339894 & $Z_5 Z_6$ & 0.0007740090482 & $X_9 X_{10}$ & 0.0002306927820 \\
  $Y_7$ &  0.0003111400198  & $Z_2 Z_3$ & 0.0000869601531 & $X_6 X_7$ & 0.0003856405199 & $X_9 Y_{10}$ & 0.0008086945582 \\
  $Z_7$ &  0.0              & $X_3 X_4$ & 0.0008090860326 & $X_6 Y_7$ & 0.0017265374157 & $X_9 Z_{10}$ & 0.0014672289588 \\
  $X_8$ &  0.0021851855014  & $X_3 Y_4$ & 0.0009637438292 & $X_6 Z_7$ & 0.0011718305575 & $Y_9 X_{10}$ & 0.0 \\
  $Y_8$ &  0.0006081573813  & $X_3 Z_4$ & 0.0008743308693 & $Y_6 X_7$ & 0.0003206445689 & $Y_9 Y_{10}$ & 0.0009904281035 \\
  $Z_8$ &  0.0010763953485  & $Y_3 X_4$ & 0.0002868450783 & $Y_6 Y_7$ & 0.0003806542922 & $Y_9 Z_{10}$ & 0.0002488933528 \\
  $X_9$ &  0.0006056925257  & $Y_3 Y_4$ & 0.0013018963990 & $Y_6 Z_7$ & 0.0009424159528 & $Z_9 X_{10}$ & 0.0003356747248 \\
  $Y_9$ &  0.0008321612132  & $Y_3 Z_4$ & 0.0000797314206 & $Z_6 X_7$ & 0.0005749526574 & $Z_9 Y_{10}$ & 0.0 \\
  $Z_9$ &  0.0010019294656  & $Z_3 X_4$ & 0.0002635981122 & $Z_6 Y_7$ & 0.0009080912917 & $Z_9 Z_{10}$ & 0.0003655776749 \\
  $X_{10}$ &  0.0           & $Z_3 Y_4$ & 0.0001441857959 & $Z_6 Z_7$ & 0.0012534980664 &  &  \\
\end{tabular}
\end{ruledtabular}
\end{table}

\end{document}